\RequirePackage{fix-cm}
\documentclass[natbib,smallextended]{svjour3}       
\smartqed  
\usepackage{graphicx}
 \journalname{my journal}
\begin{document}

\title{Solar cycle indices\\ from the photosphere to the corona:\\measurements and underlying physics
}

\titlerunning{Solar cycle indices from the photosphere to the corona}        

\author{
        Ilaria Ermolli \and
        Kiyoto Shibasaki \and \\
        Andrey Tlatov \and Lidia van Driel-Gesztelyi        
}


\institute{I. Ermolli \at
              INAF Osservatorio Astronomico di Roma, via Frascati 33, 00040 Monte Porzio Catone, Italy\\
              \email{ermolli@oaroma.inaf.it} 
              \and
              K. Shibasaki \at
              Nobeyama Solar Radio Observatory NAOJ, 462-2 Nobeyama, Minamimaki, Minamisaku, Nagano 384-1305, Japan\\
                         \email{shibasaki.kiyoto@nao.ac.jp}
              \and
              A. Tlatov \at
              Kislovodsk Mountain Astronomical Station of the Pulkovo Observatory, Kislovodsk, Russia\\
             \email{tlatov@mail.ru}   
             \and
             L. van Driel-Gesztelyi \at
              University College London, Mullard Space Science Laboratory, Holmbury St. Mary, Dorking, Surrey RH5 6NT, UK \\
            LESIA, Observatoire de Paris, CNRS, UPMC, UniversitŽ Paris Diderot, Paris, France\\
Konkoly Observatory of the Hungarian Academy of Sciences, H-1121 Budapest, Hungary\\
              \email{Lidia.vanDriel@obspm.fr}      
 }

\date{Received: date / Accepted: date}

\maketitle

\begin{abstract}
A variety of indices have been proposed in order to represent the many different observables modulated by the solar cycle. Most of these indices are highly correlated with  each other owing to their intrinsic link with the solar magnetism and the dominant eleven year cycle, but their variations may differ in fine details, as well as on short- and long-term trends.
In this paper we present an overview of the indices  that are often employed to describe the many features of the  solar cycle, moving from the ones referring to direct observations of the inner solar atmosphere, the photosphere  and chromosphere, to those deriving from measurements of the transition region and solar corona. For each index, we summarize existing measurements {\bf and typical use}, and for those that quantify  physical observables, we describe the underlying physics. 

\end{abstract}

\section{Introduction}
\label{intro}

The solar cycle, i.e. the cyclic regeneration of the appearance  of the Sun that occurs with a dominant eleven year period, originates 
below the 
 visible surface of the Sun.  In a thin shear layer  at the bottom of the convection zone, the turbulent convection operates, jointly with rotational shear, global circulations and  boundary layers, to produce magnetic fields by dynamo processes. These fields thread their way to the solar surface, where they   manifest themselves   with 
 the rich variety  of features  observed on the solar atmosphere. The   evolution of these features constitutes the solar activity cycle. Most recent models and  observations  of  the solar cycle are reviewed  e.g. by \citet{Charbonneau_2010}, \citet{Charbonneau_2014} and \citet{Hathaway_2010}.
 
Cyclic variability has been detected in cool stars like the Sun \citep{Hall_2008,Olah_etal2009,Reiners_2012,Chaplin_etal2014}. However, the Sun is the only star where its activity cycle has been studied in detail.  
Indeed, the Sun has been monitored systematically since the advent of the telescope in the early 17th century. 

 Early observations have shown that  the Sun has sunspots that  move on the solar disk due to the {\bf solar differential rotation }   \citep[e.g.][]{Galilei_1613,Scheiner_1626}. More that two centuries later
the number of sunspots was  recognized to have a cyclic variation \citep{Schwabe_1843}. 
Then, it was  found that the latitudinal distribution of sunspots and its progression over the sunspot cycle  follow a ``butterfly diagram'' \citep{Maunder_1904}.  {\bf This diagram  is now a representative image of the solar cycle.}

{\bf Sunspots were give a foundation of physics in the early decades of 20th century, when they were  demonstrated to be the seat of strong magnetic fields \citep{Hale_1908}. The magnetic laws of the solar cycle became apparent 
\citep{Hale_etal1919,Hale_1924,HaleNicholson_1925} once the measurements of  magnetic fields in sunspots  were extended over more than a single sunspot cycle.}  Sunspot and later polar magnetic  field measurements also revealed  the reversal of the global magnetic field of the Sun with the period of 22 years, the solar (or sunspot) cycle being half of the magnetic cycle \citep{Babcock_1959}. 
\begin{figure*}
\includegraphics[width=5.5cm]{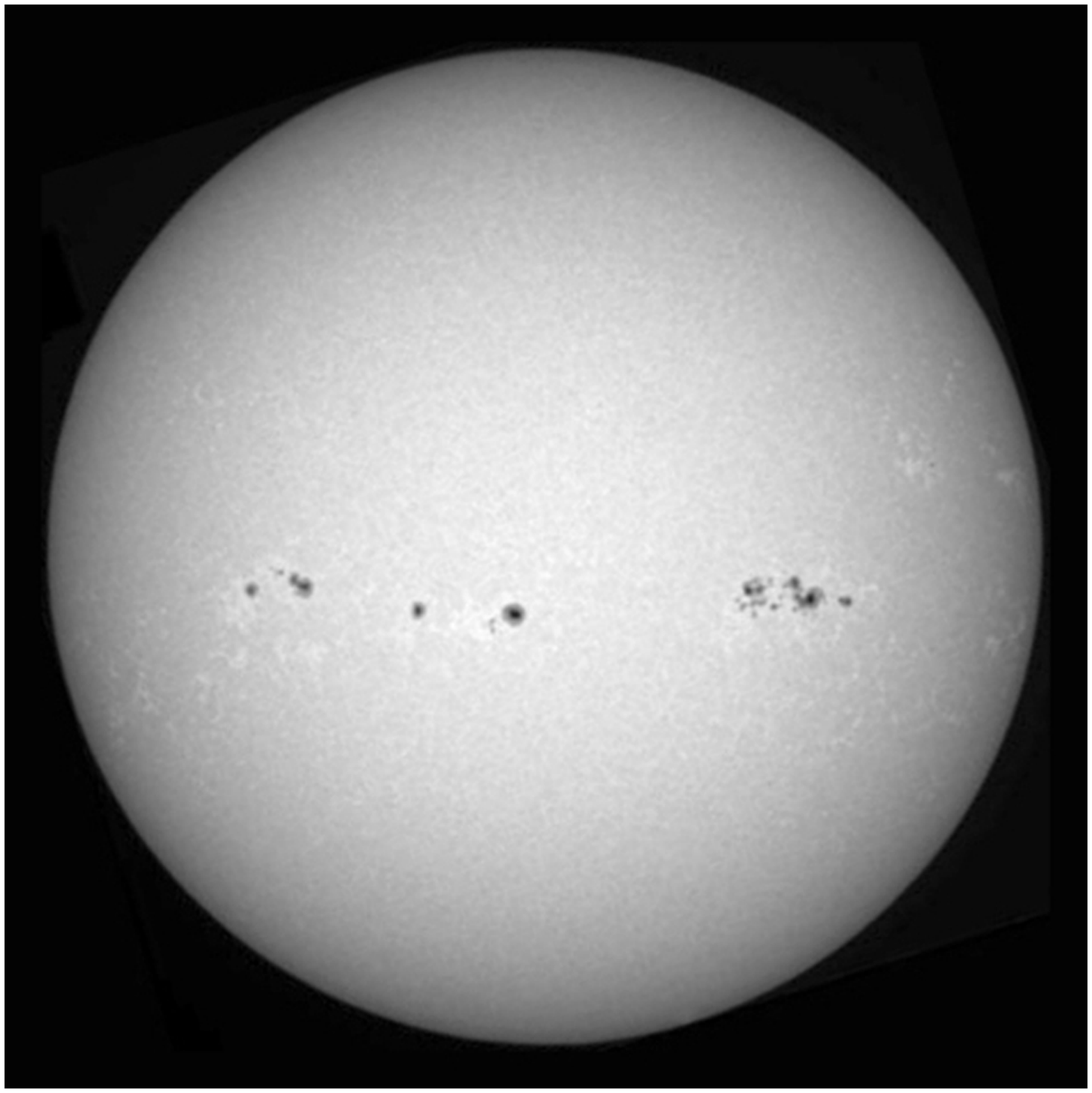}\includegraphics[width=5.5cm]{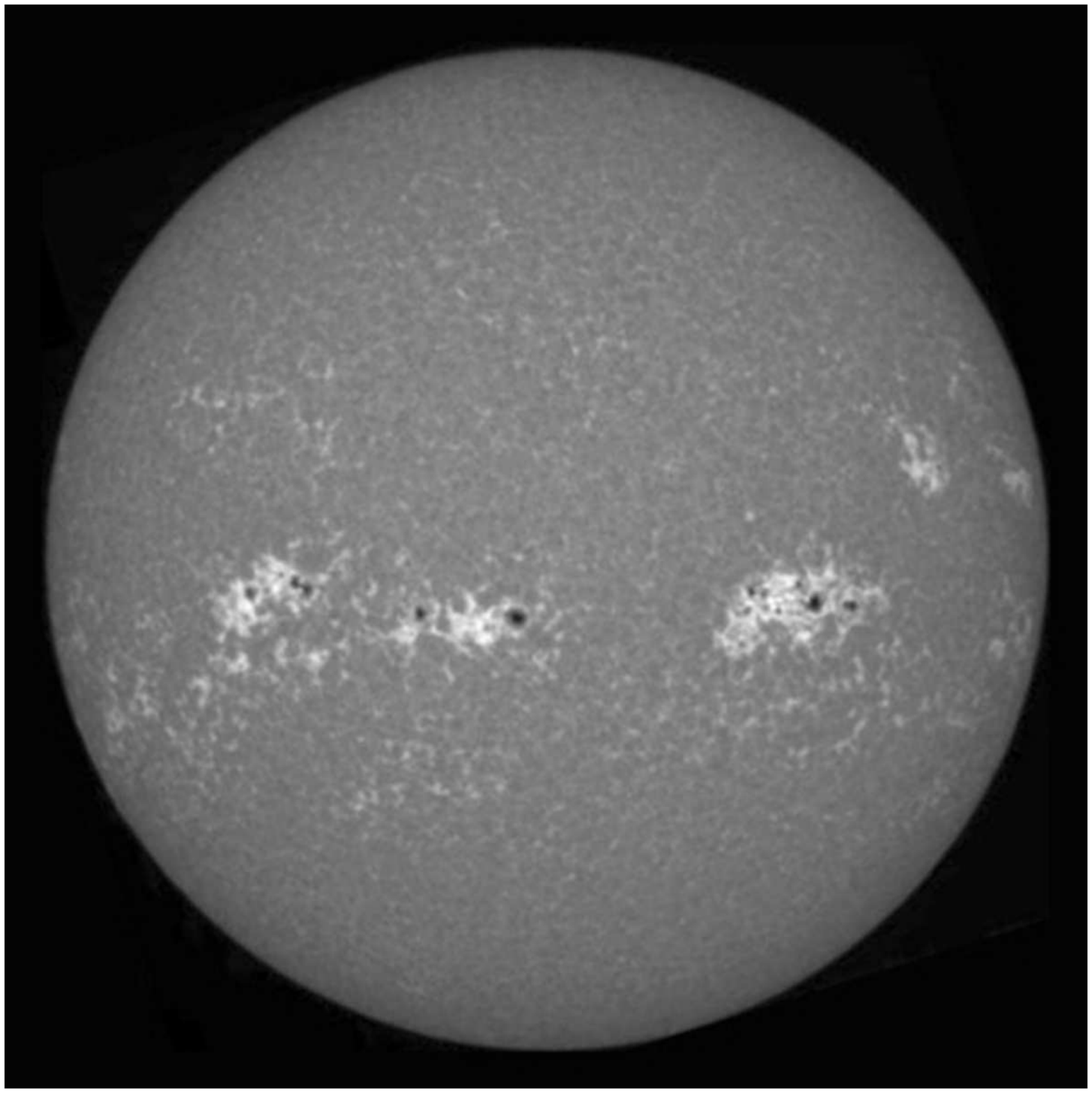}
  \includegraphics[width=5.5cm]{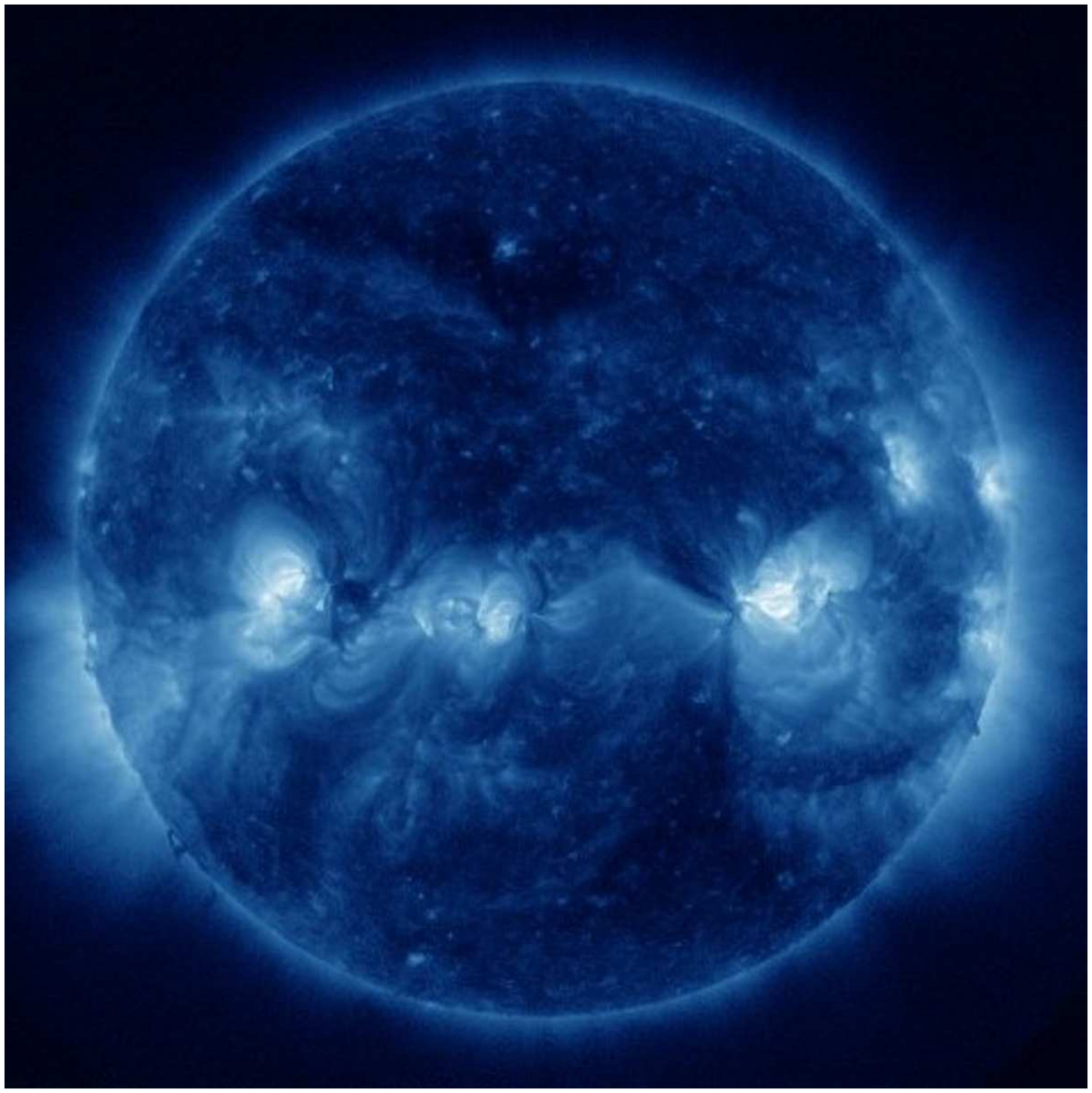}\includegraphics[width=5.5cm]{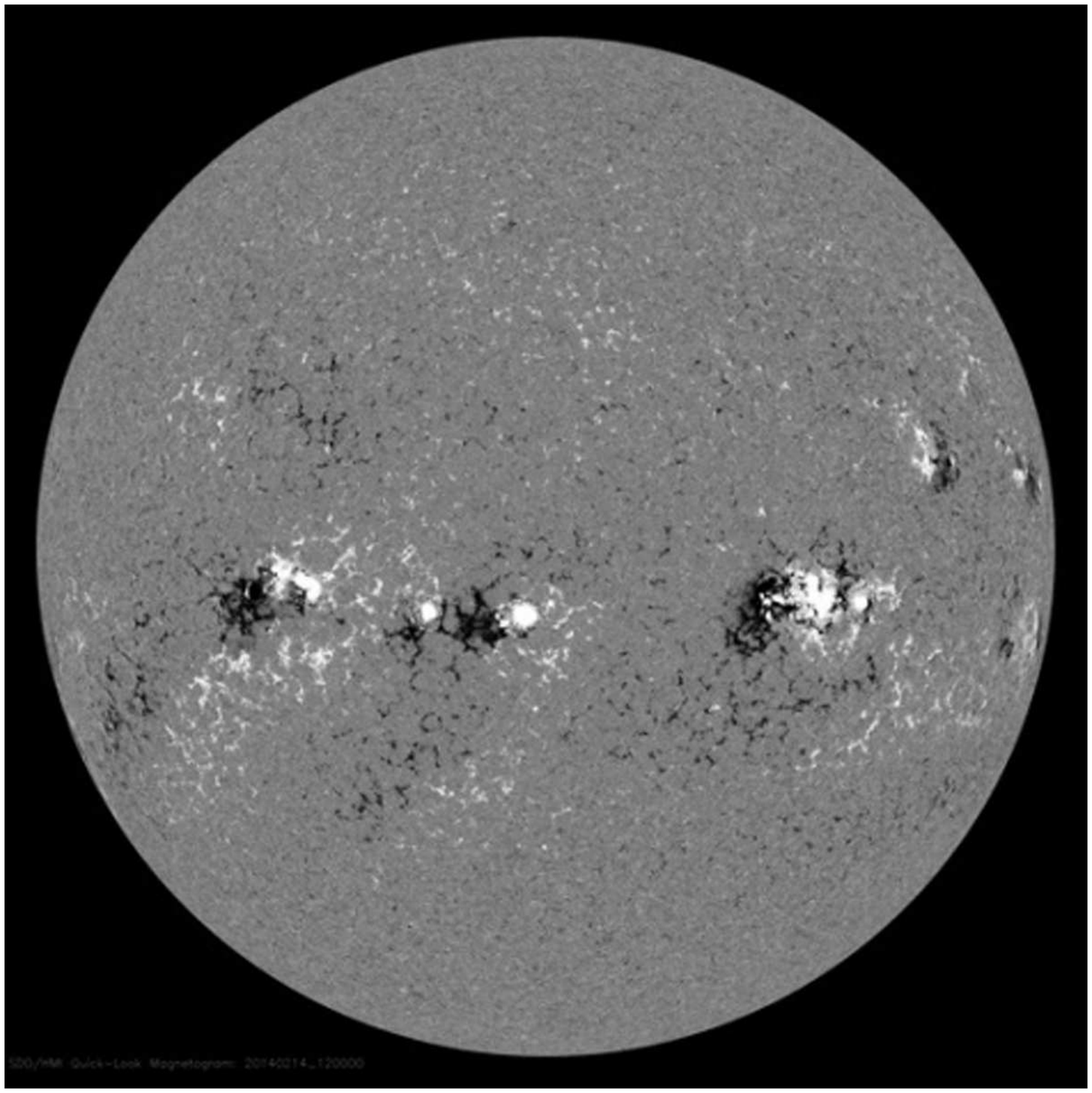}
\caption{Simultaneous views of the Sun at different heights in the solar atmosphere and measurements of magnetic field in the photosphere. Top panels: The photosphere (left) observed in the light of the wavelength 430 nm of the CH G-band on 14 February 2014, showing sunspots and faculae, the latter being apparent as bright regions mostly near the disk limb. The chromosphere  (right) observed at the  Ca II K  line radiation at wavelength 393 nm showing the enhanced chromospheric emission from the photospheric magnetic regions. The chromospheric extensions of the faculae (the so-called ``plages'') are readily visible across the whole disk (images courtesy INAF Osservatorio Astronomico di Roma). Bottom panels: The contemporaneous views of active regions and full-disk solar corona (left) observed at wavelength 33.5 nm  of the Fe XVI line    and magnetogram (right) measurement of the line-of-sight (LOS) magnetic field  (images courtesy SDO/AIA and SDO/HMI).}
\label{fig:0}       
\end{figure*}

\begin{figure*}
  \includegraphics[width=12cm]{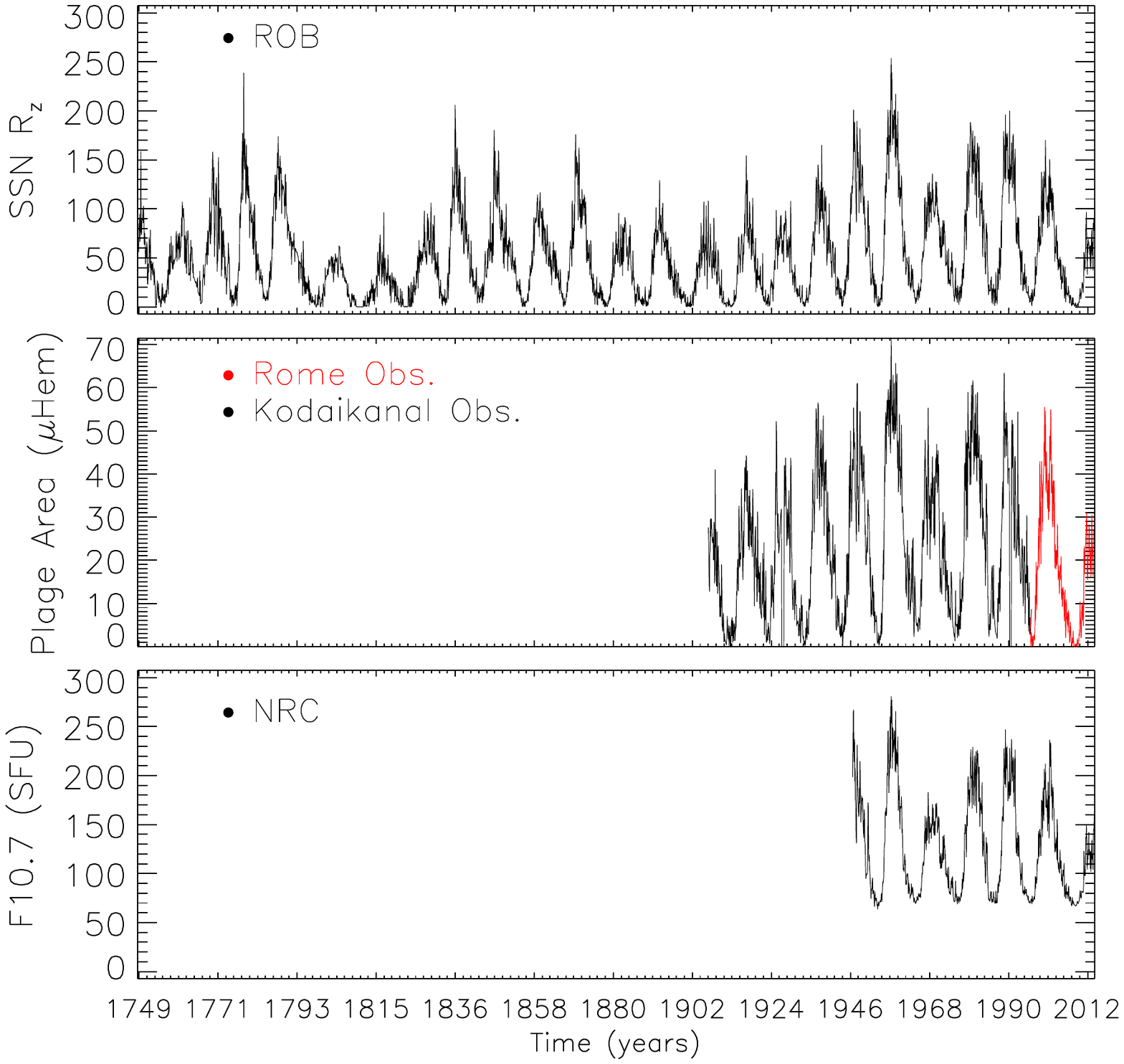}
\caption{Comparison of three indices of solar cycle.  Monthly averaged values of the international sunspot number (top panel) from the {\bf ROB-SILSO} archive (January 1749 -- March 2014),  the Ca II K plage area (middle panel) from observations carried out at the Kodaikanal (black symbols) and Rome (red symbols) observatories (February 1907 -- August 1996 and May 1996 -- November 2013, respectively), and the  F10.7 index (bottom panel) from the NRC series (February 1947 --  March 2014).  {\bf The plage area is   expressed in millionths of a solar hemisphere ($\mu$Hem), whereas the F10.7 index values are given using the Solar Flux Units (SFU, 10$^{?22}$ Wm$^{?2}$Hz$^{?1}$). } All indices show common features (eleven year cycle, long-term trends), but each index also reflects a specific aspect of solar activity.}
\label{fig:1}       
\end{figure*}

Sunspots are the most commonly known, but only  one, among the many manifestations of the solar  magnetic fields, which structure the Sun's atmosphere and make it variable \citep[e.g.][]{Solanki_etal2006,Stenflo_2008,Stenflo_2013}. This is shown in Figure  \ref{fig:0}, which presents  
three simultaneous  views of the Sun at different heights in the solar atmosphere, from the photosphere to the corona,  and the contemporaneous measurements of the magnetic field in the photosphere.

 The  cyclic  variability of sunspots was found to affect  all the manifestations of the solar magnetism on the Sun's atmosphere, including  striking  eruptive events that are driven by the evolution  of magnetic fields at small scales, such as flares, prominences, and coronal mass ejections. The number of these events  follows the rise and decay of the solar cycle \citep[e.g.][and references therein]{Hudson_etal2014}. {\bf The solar cycle also  characterizes the radiant behavior of the  Sun \citep{Willson_etal1981,Hudson_etal1982,Domingo_etal2009}
 and there is growing evidence that changes in solar irradiance affect the Earth's
middle and lower atmosphere \citep[e.g.][and references therein]{Ermolli_etal2013,Solanki_etal2013}.}
  Besides, as solar magnetic fields are dispersed into the heliosphere with the solar wind, 
the solar cycle
modulates the particulate and magnetic fluxes in the heliosphere, 
by determining  the interplanetary conditions and imposing electromagnetic forces that affect the planetary atmospheres, including the Earth's atmosphere \citep[][]{Cranmer_2009}. {\bf Indeed, the large scale structure and dynamics of the magnetic field in the heliosphere is governed by the solar wind flow, which  has its origin in the magnetic structure of the solar corona driven by the  emergence of magnetic flux \citep[e.g.][and references therein]{Schmidier_etal2014},  plasma motions in the photosphere and transient solar eruptions in the corona. Most of the flux emerged in the photosphere forms chromospheric or coronal loops that do not contribute to the heliospheric magnetic field carried by the solar wind, but a fraction of flux  extends out to form the heliospheric magnetic field, whose variations are an important source of geomagnetic activity \citep[e.g.][]{Pulkkinen_2007,Pulkkinen_etal2007}.}
Interplanetary transients and geomagnetic disturbances are found to be related to the changing magnetic fields on the solar surface\citep[][]{Lockwood_2013,Owens_2013,Wang_2014}, which indirectly modulate also the flux of high-energy galactic cosmic rays entering the solar system from elsewhere in the galaxy \citep[e.g.][]{Usoskin_2013}.  Besides, as  $^{14}$C and $^{10}$Be radioisotopes are produced in the Earth's stratosphere by the impact of galactic
cosmic rays on $^{14}$N and $^{16}$O, the solar cycle modulation of the cosmic ray flux  also leads to
solar cycle related variations in the abundances of $^{14}$C \citep{Stuiver_etal1980} and
$^{10}$Be \citep[][]{Beer_etal1990} in the Earth's atmosphere. {\bf Records of geomagnetic and cosmogenic isotope abundance variations  
can then be used to infer the near-Earth solar wind conditions and the magnetic field structure and intensity \citep[][]{Lockwood_2013,Svalgaard_2014,Hudson_2014}. \citet[][]{Lockwood_2013} gives detailed description of a large number of geomagnetic indices  linked to the amount of heliospheric open flux, allowing its reconstruction of from historic geomagnetic activity observations. }

 A variety of indices  have been proposed in order to represent the many different observables modulated by the solar cycle. Most of these indices are highly correlated with each other as they are intrinsically inter-linked through solar magnetism  and its dominant eleven year cycle. {\bf However,  they may differ in many features, as well as on  short- and  long-term trends. This is shown in 
   Figure \ref{fig:1}, which presents } the time series of monthly averaged values of three solar cycle indices discussed in the following. All measurements show the eleven year solar cycle, a long-term trend with  increasing values in time from 1900 to 1960 and then a reduced solar activity. 
Apart from these common features, these indices also show clear differences, in particular during the solar minima and on the long-term trends. This is because 
the compared  indices are related in various ways to different aspects of magnetic processes taking place on the Sun.

{\bf We  here give an overview of}  the  indices that are often employed to  represent solar cycle properties, moving from the ones resulting from direct observations of the inner solar atmosphere, the photosphere (Sect. 2) and chromosphere (Sect. 3), to those deriving from measurements of the transition region and solar corona (Sect. 4). For each index, we summarize existing measurements {\bf and typical use}, and for those that quantify a directly-measurable value of a physical observable, we describe the underlying physics, before our conclusion are given (Sect. 5). The indices discussed in this paper derive only from observations of the solar atmosphere. Most of these observations have been obtained  with ground-based instruments over many decades. A review of the indices that quantify effects caused by the solar magnetism  on the heliospheric and terrestrial environments, mostly due to the varying properties of the solar-wind  and interplanetary magnetic field,  can be found in this volume, e.g., in  \citet{Usoskin_etal2014} and \citet{Svalgaard_2014}, respectively. The solar cycle variation in the total and spectral solar irradiance is reviewed in this volume by \citet{Yeo_etal2014}.

\section{Indices from  observations of the photosphere}

\begin{figure*}
  \includegraphics[width=12cm]{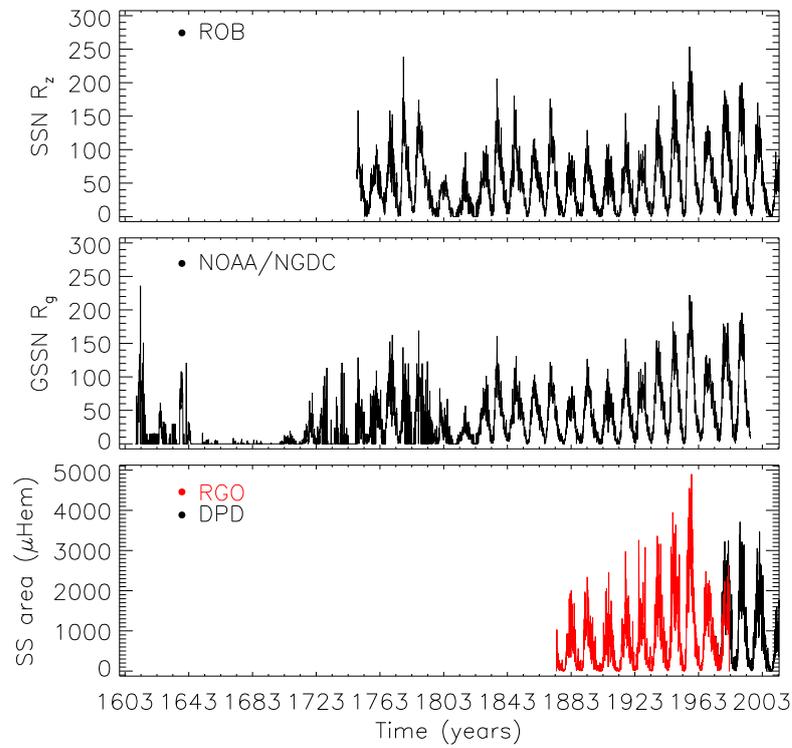}
\caption{Comparison of sunspot indices.  Monthly averaged values of the international sunspot number  (top panel) from the {\bf ROB-SILSO} archive (January 1749 -- March 2014),  group sunspot number (middle panel) from the NOAA/NGDC archive (February 1610 -- December 1995) and sunspot area (bottom panel) from the  RGO (red symbols) and DPD  (black symbols) series (January 1874 -- December 1982 and January 1974 -- April 2014, respectively).  {\bf The sunspot area is  expressed in millionths of a solar hemisphere ($\mu$Hem).}
}
\label{fig:2}       
\end{figure*}

Since the advent of the telescope, {\bf sunspots} have been monitored regularly by amateur astronomers and at many observatories. Several indices have been defined  from the observed sunspots,  {\bf e.g. the international (earlier, Z\"urich or Wolf) sunspot numbers, the group sunspot numbers, and the sunspot areas. The  time series of these indices, which are shown in  Figure \ref{fig:2} and further discussed below, are used mainly as proxies of solar activity and as tracers of the dynamo processes responsible for the build-up of large scale magnetic fields into  the solar atmosphere}. 

Apart from sunspots, faculae\footnote{Faculae is the name given to brightenings seen in photospheric radiation  mainly near the solar limb and  in the general vicinity of
sunspots. Find more information in e.g.  \citet[][]{Solanki_etal2009}.} are  the next obvious feature on the solar disk. They have been known since telescopes have been pointed at the Sun. Observations show their magnetic nature. {\bf They  represent magnetic field dispersed from strong magnetic field concentrations in sunspots. Their origin leads facular properties to vary with the sunspot cycle}.  

{\bf Regular measurements of the magnetic field strength in sunspot and faculae have been  carried out in addition to photometric observations since the advent of magnetographs. These measurements have been employed to produce line-of-sight (LOS), or radial, or vectorial estimates of the photospheric magnetic field in
physical units, and  to obtain some direct and indirect measurements of the magnetic field properties}.

\subsection{Sunspot number}
The sunspot number  indicates a weighted estimate of individual sunspots and sunspot groups derived from visual inspection of the solar photosphere in white-light\footnote{The term white-light indicates the sum of all visible wavelengths of solar radiation from 400 to 700 nm, so that all colors are blended to appear white to the eye.}  integrated radiation. The sunspot number was introduced  in the middle of the 19th century by Rudolf Wolf, who also started the program of synoptic measurements carried out at the Z\"urich Observatory that now constitutes the so-called  Z\"urich (or Wolf) sunspot number  series. {\bf This measurement series covers  the period 1849--1981, but it was extended for several solar cycles backwards by using data available from earlier observations. }

In this series, the sunspot number  $R_z$ is defined as:
\begin{equation}
R_z = k (10 G + N),
\end{equation}
where G is the number of sunspot groups, N is the number of individual sunspots in all groups visible on the solar disk and k {\bf denotes a correction factor that compensates for differences in observational techniques and instruments used by the observers.} 

The  Z\"urich sunspot number series is based on observations carried out by a single astronomer for each day, by using almost the same technique  for the whole period, but an offset due to the changes of the weighting procedure introduced in 1945--1946  and other corrections  applied for earlier data. However,  it is known that  data gaps were filled with interpolation between available measurements without note on the  series, by leading to possible errors and inhomogeneities in the  data. 
Besides, 
the  single observers have changed in time, as well as also observatory instruments and practices. {\bf These changes raise} questions about  the accuracy and integrity of the sunspot number data series available to date.  We omit discussion of this important aspect of the sunspot  data and refer the reader to the review in this volume by \citet{Clette_2014} for further information.

The  Z\"urich sunspot number data set is available online at e.g. the   archive of the National Oceanic and Atmospheric
AdministrationÕs  National Geophysical Data Center (NOAA/NGDC) \footnote{http://www.ngdc.noaa.gov/stp/solar/ssndata.html}.

Since the termination of the sunspot observations at the Z\"urich Observatory
the sunspot number series has been routinely updated as the International Sunspot Number (ISN) by the World Data Center for the production, preservation and dissemination of the international sunspot number (SILSO, Sunpot Index and Long-term Solar Observation\footnote{http://www.sidc.be/silso}) at the Royal Observatory of Belgium \citep[ROB,][]{Clette_etal2007}. The ISN series is computed using the same definition as employed for the Z\"urich series, but it represents  the  weighted average of  sunspot numbers determined by several approved observers from more than 30 countries instead of that by a single observer.  It is worth noting that sunspot numbers  provided by individual stations, e.g.  the Z\"urich series, are often 20 to 50\% higher than the ISN value, due to a multiplicative factor introduced  to reduce results of modern observations to the scale of the original series started by Rudolf Wolf. 
Find  information on  the survey and merging of sunspot catalogs in e.g. \citet{Lefevre_Clette_2014}. The ISN  data are available online at the {\bf ROB-SILSO} web site\footnote{http://www.sidc.be/silso/datafiles}.

{\bf In addition to the data presented above}  there are  series of hemispheric sunspot numbers that account for spots only in the Northern and Southern solar hemispheres. {\bf These series have been used to study e.g. the North-South asymmetry of solar activity \citep[e.g.][]{Temmer_etal2002, Temmer_etal2006,Virtanen_Mursula_2014,Norton_2014}}. The data are available at e.g. the {\bf ROB-SILSO} web page. 

\subsection{Group sunspot number}

Analysis of the Z\"urich series and of early sunspot data by Douglas Hoyt and Kenneth Schatten  lead to the definition of  the group sunspot number and creation of a time series of this index  based on all the available archival sunspot records since 1610 \citep{Hoyt_Schatten_1996,Hoyt_Schatten_1998a,Hoyt_Schatten_1998b}. 

In this series, the  group sunspot number $R_g$ is defined as:
\begin{equation}
R_g = \frac{12.08}{n} \sum\limits_{i} k_i^{\prime} G_i,
\end{equation}

\noindent where $G_i$ is the number of sunspot groups recorded by the $i$-th observer, $k^{\prime}$ is the observerÕs individual correction factor, $n$ is the number of observers for the particular day, and 12.08 is a normalization number scaling the group sunspot number to sunspot number values for the period of 1874--1976. The group sunspot number  is more robust than the sunspot number since {\bf  it is based on the easier identification of sunspot groups than  of individual sunspots. Besides, it  does not include the  number of individual sunspots that  is strongly influenced by instrumental  and observational conditions. 
Therefore, although the group sunspot number series suffers  some uncertainties  \citep[see, e.g.][]{Vaquero_etal2012,Leussu_etal2013}},  it is more reliable and homogeneous than the Z\"urich series. The two series are nearly identical after the 1880s \citep[e.g.][]{Hathaway_Wilson_2004}.

The group sunspot number data set is available online at the NOAA/NGDC archive\footnote{http://www.ngdc.noaa.gov/stp/solar/ssndata.html}.

In the recent years,  the recovery, digitization, and analysis of archived sunspot  drawings regularly produced by several observers from early decades of 17th to late 19th centuries  \citep[e.g.][and references therein]{Arlt_etal2013, Vaquero_Trigo_2014,Arlt_2014} 
has allowed improvement of earlier uncertain sunspot data  and  extension  for several cycles back in time of the butterfly diagram.

\subsection{Sunspot area}
Sunspot areas and positions were diligently recorded by the Royal Observatory in Greenwich  (RGO) from  1874 to  1976, {\bf by using } measurements from photographic plates obtained at the RGO and other observatories (Cape Town, South Africa, Kodaikanal, India, and Mauritius). Both umbral areas and whole spot areas were measured and corrected for apparent distortion due to the curvature of the solar surface \citep{Willis_etal2013a,Willis_etal2013b}.   Sunspot areas are given in units of millionths of a solar hemisphere. 

Since 1976 the RGO measurements have been continued in the  Debrecen Photoheliographic Data (DPD) sunspot catalogue {\bf that is  compiled by the Debrecen Heliophysical Observatory,  as commissioned by the International Astronomical Union}. The DPD catalogue includes  the heliographic positions and the areas of the sunspots on the full-disk white-light images obtained  at the Debrecen and Gyula observatories, as well as from other observatories.  The DPD sunspot data {\bf  are available online\footnote{http://fenyi.solarobs.unideb.hu/DPD/index.html} together with } images of sunspot groups, scans of full-disk white-light observations, and magnetic observations. Moreover,  {\bf the US Air Force  also started compiling sunspot data
from the Solar Optical Observing Network of telescopes (USAF/SOON) \footnote{http://www.ngdc.noaa.gov/stp/solar/sunspotregionsdata.html} since RGO ceased its program}. 
Measurements of sunspot area are also available from a number of solar observatories around the world e.g. Catania (1978 -- present),  Kislovodsk (1954 -- present), Kodaikanal (1906 -- present), Mt. Wilson (1917 -- 1985), Rome (1958 -- present), and Yunnan (1981 -- present).

The data recorded at the RGO and  the other observatories listed above are available at e.g. the NOAA/NGDC archive\footnote{http://www.ngdc.noaa.gov/stp/solar/sunspotregionsdata.html}. {\bf This archive includes} some fragmentary data of sunspot areas obtained for earlier periods from solar drawings  \citep[e.g.][]{Lepshokov_etal2012,Arlt_etal2013,Vaquero_Trigo_2014}.

 {\bf  Sunspot areas are considered to be a more physical characterisation of the solar cycle than sunspot or group numbers, due to the linear relation between sunspot area and total magnetic flux of the sunspot \citep[see e.g.][and references therein, and see \citet{Preminger_2007}  for the underlying cause of non-linearities]{Kiess_etal2014}. Besides, sunspot area  data have the additional information  on the disc position of the observed features  with respect to sunspot number series}.  For example, Figure    8 of \citet[][]{Hathaway_2010} shows the butterfly diagram of the 
daily sunspot area values as a function of latitude and time from 1874 to 2010 from the RGO and USAF archives.

A number of studies have attempted to inter-calibrate the sunspot area measurements available from the various  observatories. {\bf The data}  were found to be not uniform across the various sets and even within a given  set
\citep[e.g.][and references therein]{Baranyi_etal2001,Balmaceda_etal2009,Baranyi_etal2013}. For example, 
the values reported by USAF/SOON series resulted to be significantly smaller than those from RGO measurements \citep[][]{Balmaceda_etal2009}. 
  \citet[e.g.][]{Li_etal2009} found that the time series of  area values corrected for foreshortening on the solar disk and the ISN values  are  highly correlated, with a  high level of phase synchronization between the compared indices in their low-frequency components (7-12 years),  and a noisy behavior with strong phase mixing  in the high-frequency domain, mainly  around the minimum and maximum times of a cycle.

{\bf Sunspot area measurements are main input data e.g. to models of total magnetic flux and  solar irradiance variations  \citep[e.g.][and references therein]{Preminger_2006,Preminger_2007,Pagaran_etal2009,Krivova_etal2010}.}

\subsection{Magnetic field measurements}

Magnetograms are  measurements of the net magnetic field strength averaged over the resolution element of the observation and polarity at a given height in the  solar atmosphere. 
The instruments employed to produce these data rely on a variety of techniques applied to measure 
the polarization of light at various wavelength positions within a solar spectral line. Circular polarization in the opposite sense on either side of a magnetically sensitive spectral line gives a measure of the longitudinal component of the magnetic field vector. Linear polarization provides information on the strength and direction of the magnetic field transverse to the LOS.
These measurements have been employed to produce  LOS, or radial, or vectorial estimates of the observed magnetic flux density in
physical units (either Tesla or Gauss, by using SI or cgs units, respectively).

Full-disk solar magnetograms have been recorded on a daily basis, starting
at the Mt Wilson Observatory  \footnote{http://obs.astro.ucla.edu/intro.html} in the late 1950s, then with higher spatial resolution
at the National Solar Observatory (NSO) Kitt Peak  since  the early 1970s, with the Kitt Peak Vacuum Telescope  (KPVT, 1974 -- 2003)\footnote{http://diglib.nso.edu/ftp.html}  and  
  since 2003  with the Synoptic Optical Long-term Investigations of
the Sun (SOLIS) vector-spectromagnetograph  (VSM, 2003 -- present)\footnote{http://solis.nso.edu/0/index.html}, at the Wilcox Solar Observatory   (WSO, 1975 -- present)\footnote{http://wso.stanford.edu}, and more recently by instruments of the Global Oscillations Netowrk Group (GONG, 2006 -- present) \footnote{http://gong.nso.edu}. The measurements have been also performed in space with the
SOHO Michelson Doppler Imager  magnetograph (SOHO/MDI, 1996 -- 2011) \footnote{http://soi.stanford.edu/data/}, and  presently 
by SDO Helioseismic and Magnetic
Imager  (SDO/HMI, 2010 -- present)\footnote{http://jsoc.stanford.edu}. The first KPVT observations are characterized by  4 arcsec resolution and daily cadence, while the latter data obtained with the SDO/HMI are characterized by 1 arcsec resolution and 45 second cadence.  Most of the magnetogram time series listed above  refer to photospheric observations of the either the LOS  or radial photospheric magnetic field strength, but the recent SOLIS observations  {\bf that are taken by sampling two spectral lines originating} at  photospheric and chromospheric heights.
SOLIS and SDO/HMI observations also produce  vector magnetic field measurements.  For example, the synoptic magnetic field movie linked to Figure  13 of 
\citet{Hathaway_2010} shows a magnificent evolution of the solar magnetism during the last  two and half solar cycles using NSO magnetograms between 1980 -- 2009.

The results of magnetic field  measurements are available at the web page of the various institutes and space missions listed above. An online  database at the Pulkovo Observatory\footnote{http://www.gao.spb.ru/database/mfbase/main\_e.html} 
contains data of magnetic field measurements obtained since 1957 at a number of  observatories  of the former Soviet Union.

{\bf The field measurements  have been stored at the various observatories as are and  further processed  to produce synoptic charts. These charts}  show   the   latitude-longitude distribution of the magnetic field during a complete solar rotation and the butterfly diagram patterns. For example, Figure  1 of \citet[][]{Petrie_2013} shows the butterfly diagram obtained from NSO Kitt Peak data, summing up the photospheric radial field distributions derived from the longitudinal photospheric field
measurements from 1974 to 2013. {\bf These charts are the main input data  to numerical models of e.g. the solar dynamo and of the solar outer atmosphere and inner heliosphere \citep[e.g.][]{Riley_etal2011}}.

The results of the synoptic measurements carried out at the WSO\footnote{http://wso.stanford.edu}, NSO, and other observatories have been also {\bf employed to derive the time series of  the monopole, dipole, and higher order coefficients of the spherical harmonics}  that  decompose the  magnetic field observations as a function of the  surface latitude and longitude. The multipole coefficients   {\bf have been  used to study the dominant length scales and symmetries of the observed field \citep[e.g.][]{Petrie_2013}, as well as to predict the value of the sunspot numbers at the next solar maximum \citep[see e.g.][]{Pesnell_2012}}.  {\bf The heliospheric consequences of the solar cycle variation of the Sun's low-order magnetic multipoles are discussed in this volume by e.g. \citet{Wang_2014}.}

Measurements carried out at the Mt Wilson Observatory have been also analyzed to produce proxies of sunspot and plage properties, specifically the {\bf Mount Wilson Sunspot and Magnetic Plage Strength Indices, which  represent  the fractional solar
surface covered by  magnetic fields
exceeding 100 Gauss and 
 between 10 and 100 Gauss, respectively}. {\bf These indices have been employed e.g. to model solar irradiance variations \citep[][]{Jain_etal2004}.}
Moreover, local helioseismology methods have provided a way to map
medium-to-large magnetic regions on the far-side hemisphere of the Sun \citep{Lindsey_Braun1997,Gonzales_etal2007}. The sound waves that pass
through areas of strong magnetic fields experience a phase shift (Braun et al. 1992) that can be estimated with respect to the propagation in a non-magnetized atmosphere. 
Maps of the difference between the model and the measured
phase shift  have been derived  regularly {\bf during the last decade } from 24 hours of solar  surface velocity data obtained by  the GONG\footnote{http://farside.nso.edu}  and SOHO/MDI observations. These maps  show  the
 locations of shorter travel times due to change of sound speed in strong magnetic-field regions, whose presence modify the local plasma temperature.  SDO/HMI\footnote{http://stereo-ssc.nascom.nasa.gov/beacon/beacon\_farside.shtml}  is providing data continuation 
after the decommission of SOHO/MDI. Recent improvements {\bf have turned the far-side maps into a useful space weather forecasting tool}   \citep{Gonzales_etal2014}.

A number of studies have attempted to inter-calibrate the LOS or radial magnetograph full-disk measurements from one or more observatories (e.g. Liu et al., 2012, and references therein). 
 \citet{Riley_etal2014} compared maps from several  of the observatories listed above to identify consistencies and differences among them. They found that while there is a general qualitative agreement among the maps produced by the various programs, there are also some significant differences. They also computed  conversion factors that relate measurements made by one observatory to another using both synoptic map pixel-by-pixel and histogram-equating techniques. {\bf Besides, they  explored the relationship between the various data sets over more than a solar cycle, by finding that  the conversion factors remain relatively constant for most of the studied series}.

\subsection{White-light facular indices}
\begin{figure*}
  \includegraphics[trim=0.2cm 1cm 0.2cm 1cm,clip=true,width=8cm,angle=90]{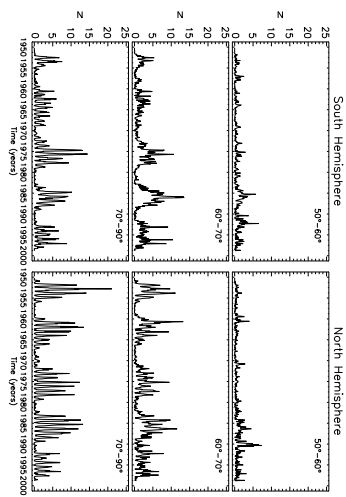}
\caption{Monthly averaged value of the  number of white-light polar faculae   measured from January 1951 to December 1998  at three high-latitude bands:   50--60$^{\circ}$ (top panels), 60--70$^{\circ}$ (middle panels), and 70--90$^{\circ}$ (bottom panels) on the South (left panels) and North (right panels) solar hemispheres. These measurements are from the  archive of the National Astronomical Observatory of Japan. 
}
\label{fig:4}       
\end{figure*}

White-light faculae were routinely measured at the RGO from 1874 to 1976, {\bf by producing a record of over 100 years of continuous daily measurements of the position and projected area of these features.} 
These  measurements are  available at the NOAA/NGDC archive\footnote{http://www.ngdc.noaa.gov/stp/solar/wfaculae.html}. 
Besides, 
polar faculae, i.e. faculae seen at high latitudes (50--90$^{\circ}$) in white-light observations, have been also monitored since late 19th century.  Regular observations of these features  have been carried out  at e.g. the RGO (1880--1954),  Mt Wilson (1906--1990), Z\"urich (1945--1965), National Astronomical Observatory of Japan (1951--1997)\footnote{http://solarwww.mtk.nao.ac.jp/solar/faculae/}, and Kislovodsk  (1960 -- present)\footnote{http://158.250.29.123:8000/web/P\_fac/} observatories. These  observations have shown the counter phase variation of the number of polar faculae with respect to the sunspot cycle. 
 Most recent studies indicate that the phase relationship between numbers of polar faculae and sunspots is both time- and  frequency- dependent \citep{Deng_etal2013}. {\bf However, the number  of faculae at the poles of the Sun appeared to be   well correlated with the LOS component of the polar magnetic field measured at the WSO \citep{Sheeley_2008}  and by the SOHO/MDI  \citep{Munoz-Jaramillo_etal2012},  suggesting that  the polar faculae number  is a good proxy to study the evolution of the polar magnetic field. Observations and models of the solar polar fields are discussed in this volume by e.g. \citet{Petrie_2014}.}

 Figure \ref{fig:4} shows 
 the monthly averaged values of the number of polar faculae  at three high-latitude bands on both  solar hemispheres from the archive of the National Astronomical Observatory of Japan. 

\section{Indices from  observations of the chromosphere}
\label{sec:1}

{\bf Monitoring programs of full-disk observations  in the Ca II  K
and H$_{\alpha}$ resonance lines have been carried out at various observatories  since the start of the 20th century. }

The Ca II  K  line at  393.37 nm is among the strongest and broadest lines in the visible solar spectrum and thus easily accessible to ground-based observations, though in the violet part of the solar spectrum. In solar observations, the core of this line shows an emission reversal, with a central absorption minimum.  In standard notation, K$_3$, K$_{2V,2R}$, and K$_{1V,1R}$ mark the core, the reversal (emission
peaks), and the secondary minima of the doubly reversed profile
of the line, in the violet (V) and the red (R) wings
of the line, respectively. All these line features occur within a
spectral range less than 1 \AA ~wide. They  result from a complex formation of the line \citep[][]{Uitenbroek_1989}, which originates over  atmospheric heights ranging from the  temperature minimum photospheric region in the line wing up to the high chromosphere in the core \citep[e.g.][]{Leenaarts_etal2013}.

Observations at the Ca II K have long served as diagnostics
of the solar chromosphere \citep[e.g.][and references therein]{Rutten_2007}. They show that the Ca II K line becomes brighter with non-spot magnetic flux. Quantitatively, the line core excess flux density, $\Delta $F$_{Ca~II}$ $(erg ~cm^{-2}s^{-1})$, can be written as:
\begin{equation}
log \Delta F_{Ca~II} = 0.6 log <B> + 4.8,
\end{equation}
where $<B>$ indicates the spatially averaged magnetic flux density in the resolution element of the observation \citep{Schrijver_etal1989}.
Therefore, {\bf the Ca II K line emission can be used as a good proxy of the LOS magnetic flux density over the whole solar disk} \citep[e.g.][and references therein]{Ermolli_etal2010}.

The H$_{\alpha}$ line at 656.28 nm has been also widely employed for studying the solar chromosphere, which was defined as what is seen in this line \citep[][]{Lockyer_1868}.
The formation of this deep-red line is complicated \citep[][]{Leenaarts_etal2012}
and  originates over   atmospheric heights ranging from the photosphere in the line wing up to the middle  chromosphere in the core \citep[e.g.][]{Leenaarts_etal2013}.   Observations at this line show a wealth of magnetically dominated solar features and processes, e.g. flares, filaments, prominences, varying pattern of large-scale magnetic polarity, and the fine structure of magnetic regions. The  appearance of these solar features depend on the  sensitivity of the H$_{\alpha}$ opacity to the mass density and temperature  of the plasma traced by magnetic fields at different heights into the solar atmosphere. 

{\bf Regular measurements of the solar  Mg II h and k resonance lines 
at 280.27 and 279.55  nm have also  been registered from space, } as a measure of solar chromospheric variability and good proxies of the solar UV emission \citep[e.g.][]{Dudok_etal2009}. These lines sample higher layers of
the solar chromosphere than the other diagnostics listed above \citep[e.g.][and references therein]{Leenaarts_etal2013}.
{\bf
The  Mg II core-to wing index  
has been derived from the Mg II measurements, } by taking the ratio of the h and k lines of
the solar Mg II emission at 280 nm to the background
solar continuum near 280 nm \citep{Heat_etal1986}. The solar Mg II index 
has been  monitored on a daily basis by NOAA
spacecraft since 1978.  
The data are available at the NOAA/NGDC archive\footnote{ftp://ftp.ngdc.noaa.gov/STP/SOLAR\_DATA/SOLAR\_UV/NOAAMgII.dat}. 

{\bf The Mg II index has been widely employed e.g. to parameterize the facular and network contribution to short- and long-term solar irradiance variations \citep{Pagaran_etal2009, Lean_etal2011,Froehlich_2013}.}

\subsection{Ca II K line and plage area indices}
\begin{figure*}
  \includegraphics[width=12cm]{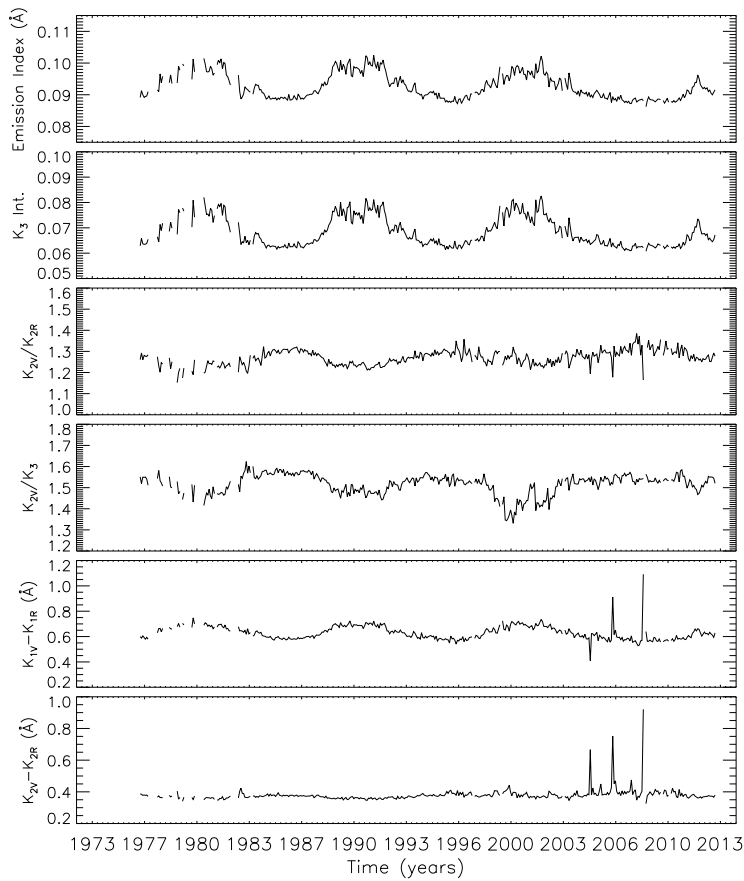}
\caption{Time series of monthly averaged values of the Ca II K line parameters measured from November 1976 to April 2014 at the NSO. Details are given in the text.
}
\label{fig:5}       
\end{figure*}

Daily full-disk Ca II K observations have been  obtained with either   spectrographs, Lyot-type, or interference filters at various observatories, e.g.  Kodaikanal (1904 -- present), Meudon (1909 -- present)\footnote{http://bass2000.obspm.fr/}, Mt Wilson  (1915 -- 1985)\footnote{http://www.astro.ucla.edu/~ulrich/MW\_SPADP}, Arcetri (1934 -- 1975)\footnote{http://www.oa-roma.inaf.it/solare/index.html}, Big Bear (1942 -- 1987), Kislovodsk  (1957 -- present)\footnote{http://old.solarstation.ru/}, Rome (1964 -- present)\footnote{http://www.oaroma.inaf.it/solare/index.html}, San Fernando (1984  -- present) \footnote{http://www.csun.edu/SanFernandoObservatory/}, 
 Kanzelh\"ohe (2010 -- present)\footnote{http://cesar.kso.ac.at}. The observations at the various sites have been reduced to measure e.g. the position and area of  Ca II K plage\footnote{Plage is the name given to the brightening seen in chromospheric radiation corresponding to photospheric faculae. 
In contrast to faculae, plage are seen over the whole disk, in active regions and in the quiet sun, on the network pattern formed at the borders of supergranular cells. Find more information in e.g.  \citet[][]{Solanki_etal2009}.} regions,  or to derive measurements of the Ca II K line indices described in the following. 
The data have been also {\bf  employed to produce synoptic charts of the  Ca II K enhanced emission, which reflects  the magnetic flux distribution on  the solar disk during a complete solar rotation.}

Measurements on modern and historical full-disk Ca II K observations show both cyclic and long-term variations of plage properties \citep[e.g.][and references therein]{Ermolli_etal2009a,Ermolli_etal2009b,Foukal_etal2009,Chapman_etal2011,Priyal_etal2014}. However, similarly to sunspot number series, caution is needed
when considering results derived from analysis of long time series of Ca II K full-disk observations,  without careful analysis of their problems and intrinsic
instrumental variations. 
In fact,  simple visual inspection of the  data available at the various archives reveals considerable differences between the images from the various  time series, due to different observational and instrumental characteristics. 
For example, \citet{Ermolli_etal2009a} show that 
the yearly median values of the plage area measurements derived from the Mt Wilson, Kodaikanal, and Arcetri time series of historical Ca II K observations agree within 40\%, with a Pearson correlation coefficient ranging from 0.85 to 0.93. However, the values derived  from the three series differ considerably for cycles 15, 17, 19, with a relative difference as high as 140\%. Besides, the time series of measured plage area values derived from the three archives show the eleven-year solar cycle variation of the  sunspot numbers, but differences at short time scales are found when comparing sunspot and plage measurements, in particular,  subtle but systematic differences around the minima of solar activity, from 1945  to the present. Middle panel of Figure  \ref{fig:1} shows the time series of monthly averaged values of the plage area derived from historical and modern Ca II K full-disk observations carried out at the Kodaikanal and Rome observatories, respectively. 
 
Some  Ca II K observations  listed above {\bf  have been  also processed to produce time series of  indices based on the intensity distribution measured over the solar disk } \citep[e.g.][]{Caccin_etal1998,Bertello_etal2010}. These indices resulted to be well correlated  to the fractional area of the solar disk occupied by plages and network. Besides, 
{\bf the full-disk Ca II K observations
have been employed to study e.g. the onset and end of solar cycles and cycle  properties
} \citep[e.g.][]{Harvey_1992,Ermolli_etal2009b}. In a recent study  \citet{Sheeley_etal2011} computed 
butterfly diagrams of the longitudinally averaged Ca II K intensity from the Mt Wilson observations for the years 1915-1985.  {\bf From analysis of these charts they found that  
cycle 19 is remarkable for 
}its broad latitudinal distribution of active regions, 
its giant poleward surges of flux, 
and for the emergence of a North-South asymmetry that lasted 10-years.
It is wort noting that the
CaII K observations exist also for prior sunspot cycles when magnetograms were not available.

{\bf Since late 1960s Ca II K line profiles of the Sun  were obtained both integrated over the solar disk and at given latitudinal bands at  e.g. the NSO  in Sacramento Peak and Kitt Peak, and at the Kodaikanal Observatory, by using high resolution spectrographs}. For the subsequent  decades, these K-line monitoring programs  have produced almost daily measurements of several  parameters
characterizing the  Ca II K-line measurements. 
{\bf The daily observations of the disk integrated Ca II K emission at the NSO  Kitt Peak  have been obtained with the  Integrated Sunlight
Spectrometer (ISS) of the SOLIS telescope since 2006.
} The ISS  monitors the solar emission at nine  different wavelength bands regularly\footnote{http://solis.nso.edu/iss}.

The  Ca II K line parameters measured to date  include the Ca K emission index, which is  defined as the equivalent width of a 1 \AA ~ band centered on the K
line core, and various measures
of line features (e.g. relative strength of the blue K$_2$ emission
peak with respect to the K$_3$ intensity,   separation of the two
emission maxima,   separation of the blue and red K$_1$ minima)  and asymmetry (ratio of the blue and red K$_2$
emission maxima). The data derived from the NSO program are available online\footnote{http://nsosp.nso.edu/cak\_mon/}. 

Figure  \ref{fig:5} shows the time series of monthly averaged values of the Ca II K line parameters measured at the NSO. 
{\bf A recent analysis of these data by \citet{Scargle_etal2013} indicates  that the temporal variation of the measured line parameters consists of five components, including the solar cycle eleven year period, a quasi periodic
variation with  100 day period, a broad band stochastic
process, a rotational modulation, and
random observational errors.  
\citet{Pevtsov_etal2013} found a weak dependency of intensity in the Ca II K line core measured in the quiet chromosphere with the phase of the solar cycle. This dependency has been attributed to  the signature  of changes in thermal properties of basal chromosphere with the solar cycle. 
}

\subsection{H$_{\alpha}$  filament and prominence data}
\begin{figure*}
  \includegraphics[trim=2cm 2cm 2cm 2cm, clip=true,width=7.5cm, angle=270]{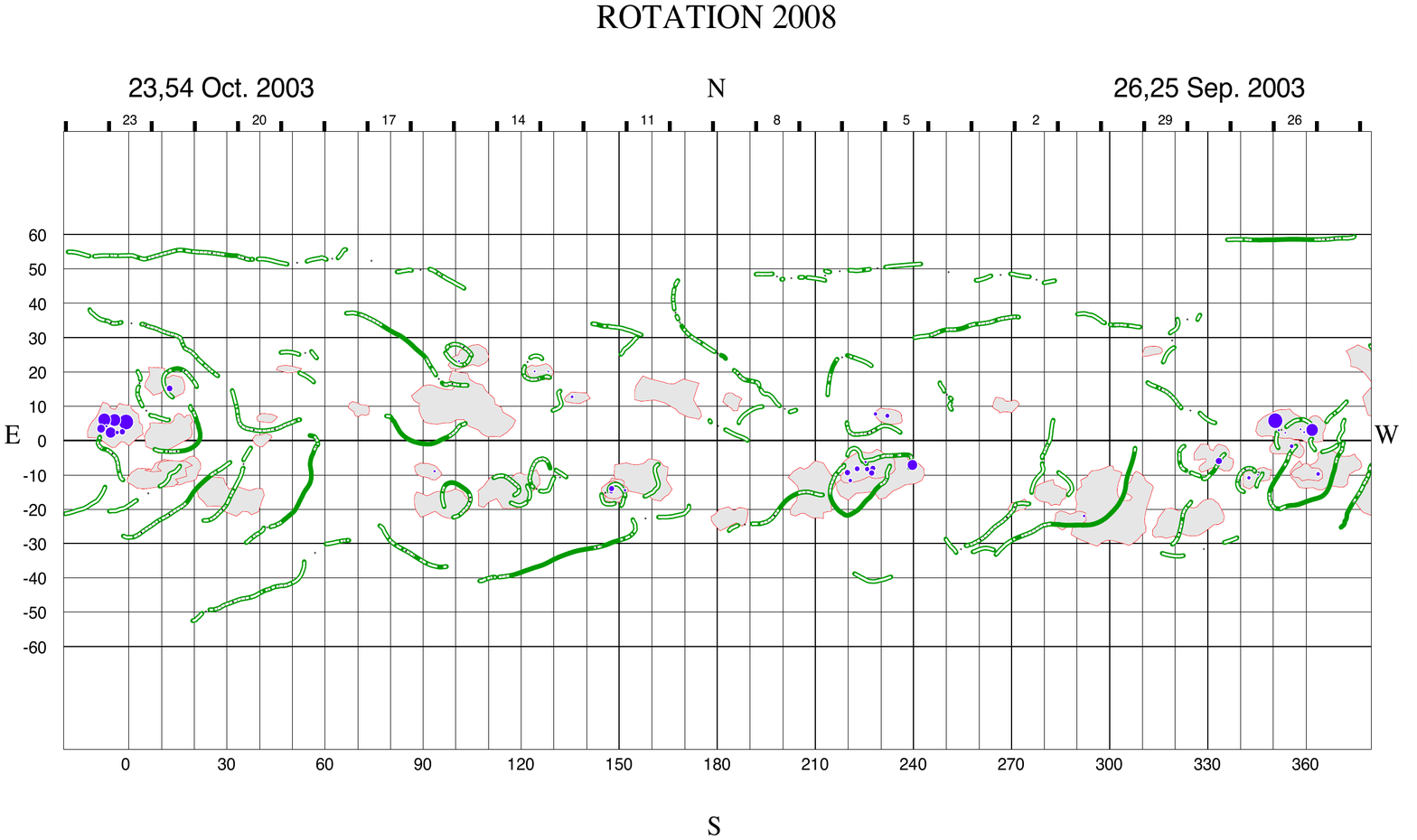}
\caption{Synoptic map of solar activity from H$_{\alpha}$ and Ca II K observations obtained at the Meudon Observatory. This map  summarizes the  properties of solar  features observed during their transit across the solar disk  from 26 September to 23 October 2003, i.e. for a solar rotation during the last activity maximum. The active  regions (sunspots and plage marked  with  blue and red areas, respectively)  and  filaments (green areas) observed on the disk are shown  at their activity maximum and  maximum spread, respectively. The position on the map of the observed features  represents the average position of daily baselines.
}
\label{fig:6}       
\end{figure*}

\begin{figure*}
  \includegraphics[width=12cm]{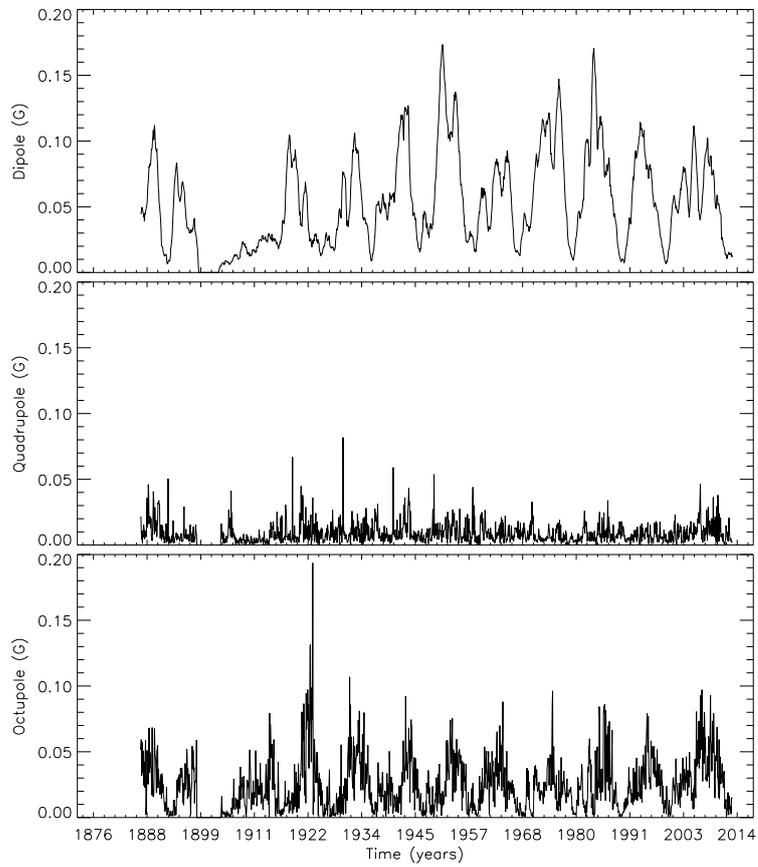}
\caption{Monthly averaged values of the dipole, quadrupole, and octupole  coefficients of the spherical harmonics {\bf that  describe the large scale solar magnetic field   from the measurements of the filaments, filament channels, and prominences in the H$_{\alpha}$ synoptic charts (January 1887 -- December 2013) produced at the Kodaikanal, Meudon, NSO Sacramento Peak,  and Kislovodsk observatories.}
}
\label{fig:3}       
\end{figure*}

Daily full-disk H$_{\alpha}$ observations have been  obtained with either  filters or spectroheliographs at  various observatories, e.g.    Meudon (1909 -- present)\footnote{http://bass2000.obspm.fr/home.php}, Arcetri (1927 -- 1969), Catania (1906 -- present)\footnote{http://www.oact.inaf.it}, Kanzelh\"ohe (1973 -- present)\footnote{http://cesar.kso.ac.at/synoptic/ha4m\_years.php}, and Big Bear (1982 -- present)\footnote{http://www.bbso.njit.edu/Research/FDHA/}, to mention a few. {\bf These observations  were taken regularly also to  support solar-terrestrial prediction services.
}

The  data  obtained at various observatories have been stored as are and further processed to produce  H$_{\alpha}$ synoptic charts, which 
incorporates
information about evolving features observed in the solar chromosphere at the H$_{\alpha}$ radiation into a single map  for each rotation
of the Sun. 
These charts show e.g.
 the boundaries separating positive and negative magnetic polarities on the solar surface, and data of the  filaments and filament channels.  At present, the summary series of these  charts cover the period from 1887 up to  present \citep[e.g.][]{Vasileva_etal2002}, though earlier data are rather uncertain. 

Figure \ref{fig:6} shows the H$_{\alpha}$ map for a solar rotation at the cycle 23 maximum, from  26 September to 23 October 2003, obtained at the 
 Meudon Observatory. 
Local H$_{\alpha}$ and Ca II K observations have been processed    \citep{Mouradian_1998}  since late 1989 for  drawing maps and for computing active region and filament tables. The produced  maps show the  properties of observed features during their transit across the solar disk. {\bf They display active regions (sunspots and plage)  at their activity maximum, i.e. maximum  area, and 
filaments  at their maximum spread at the  average position of daily baselines.} 
Other synoptic maps derived from Meudon observations show daily information of  filament traces and  prominence positions observed on the limb. Besides,  
there are data of filaments available at the NOAA/NGDC archive\footnote{http://www.ngdc.noaa.gov/stp/space-weather/solar-data/solar-features/prominences-filaments/filaments/},  dating back 1919 and consisting of the position, shape, and life  time of the observed features and of their different parts.  Other  available data  include the  area of prominences (1957 -- present) and filaments (1959 -- present)   observed  at the Kislovodsk  station of the Pulkovo Observatory.

{\bf  Figure \ref{fig:3} shows the coefficient of the spherical harmonic decomposition of the  large scale solar magnetic field derived from the field polarity data  deduced from the H$_{\alpha}$ synoptic charts produced at the  Kodaikanal, Meudon, NSO Sacramento Peak,  and Kislovodsk observatories \citep[see e.g.][for a description of the methods applied]{Makarov_2000,Makarov_2001}. It is noteworthy  that the period covered by these charts is comparable to the
length of sunspot group series, exceeding the period covered by  direct measurements of the  large-scale solar magnetic field.
}

\subsection{Flare indices} 
{\bf Flares\footnote{Flare is the name given to a sudden, rapid, and intense brightening observed over the solar disk or at the solar limb, due to a release of magnetic energy (up to 10$^{32}$ erg on the timescale of hours), followed by ejection of solar plasma  through the corona into the heliosphere. Find more information e.g. in \citet{Benz_2008}.}  are complex multi scale phenomena \citep[][]{Shibata_etal2011} that affect different spatial and temporal scales  at various heights in the solar atmosphere, from the photosphere to the corona and beyond. Therefore, they are observed at all wavelengths from decameter radio waves to gamma-rays at 100 MeV. Occasionally, flares are also seen in white-light photospheric  observations  \citep[][]{Benz_2008}. 
The X-ray and particle flux emitted during solar flares have been measured regularly since the advent of space missions.
However, before the space age, flares were  monitored for many years via  H$_{\alpha}$ chromospheric observations carried out at many observatories,  by visual, photographic, or digital  inspection of the solar disk. Although  the full effect of the  solar cycle  on flares is depicted by the measurements of these events recorded at the various wavelength bands, we focus in the following on  the flare indices from synoptic observations of the solar chromosphere, because of the remarkable archival data. Indeed, 
	reports of the H$_{\alpha}$ solar flare patrol programs } are  available from 1938 to present at the NOAA/NGDC archive \footnote{http://www.ngdc.noaa.gov/stp/space-weather/solar-data/solar-features/solar-flares/h-alpha/reports/}, which holds data for about 80 observing stations. Currently five observatories send their data to the NOAA/NGDC  archive on a routine monthly  basis.
The information stored in these reports include time of beginning-, maximum brightness-, secondary maxima-, and end- of the event, heliographic coordinates of center of gravity of flare at maximum brightness, optical brightness, area of the flaring region at time of maximum brightness. These data are   integrated
by the X-ray flux of  the events as measured from the NOAA satellites SOLRAD
(1968 -- 1974) and GOES  (1975 -- present). For GOES X-ray events, the flare is assumed to start when four consecutive 1-minute X-ray flux measurements are above a threshold value, are strictly increasing, and the last value is at least  1.4 times larger than the value measured three minutes earlier.  The maximum flux value measured during the flare defines  the  event size, which is classified by the C-, M-, X- class scale. The event ends when the flux reading returns to half the  sum of the flux at maximum plus the flux value at the start of the event.

The  flare activity shown by the H$_{\alpha}$ observations has been also described  with the  ``flare index'', whose concept was introduced by  
 Josip Kleczek in early 1950s \citep[][]{Kleczek_1952}. {\bf This quantity  is defined as the product of the intensity scale of the flare observed in the H$_{\alpha}$ radiation and its
duration in minutes. It  is assumed to be roughly proportional to the total energy emitted
by the flare  \citep[e.g.][and references therein]{Ozguc_etal2003}. } Daily data of the flare index are available at  e.g.  the NOAA/NGDC 
 archive\footnote{ftp://ftp.ngdc.noaa.gov/STP/SOLAR\_DATA/SOLAR\_FLARES/INDEX} and, for solar cycles 20 to 23, also at the site of the   Bogazici University\footnote{ftp://ftp.koeri.boun.edu.tr/pub/astronomy/flare\_index}.
 
Alternative definitions of the flare index exist with respect to that presented above,  e.g. the ``X-ray flare index'' as described e.g. by \citet[][]{Criscuoli_etal2009}, based on X-ray flux measurements in the 1-8 \AA ~range of the solar spectrum, and  the ``Comprehensive Flare Index'' computed by Helen Dodson and Ruth Hedeman from observations of the McMath-Hulbert Solar Observatory for the period from 1955 to 2008. This latter index is   based on H$_{\alpha}$ observations and measurements of the magnitude of the event measured at radio frequencies, as well as of the
importance of sudden ionospheric disturbance. The data are  available at the NOAA/NGDC archive. 

{\bf The flare index has been widely employed to single out signatures of the flaring activity in magnetic regions. For example, it has been recently used to test the efficiency of  measurements of the topological complexity of the magnetic field concentrations  to discriminate between flaring and non-flaring regions \citep[e.g.][and references therein]{Georgoulis_2013,Ermolli_etal2014}.}


\section{Indices from  observations of the transition region and solar corona}
       
{\bf 
The solar corona has been observed regularly since the advent of the coronographs by Bernard Lyot in the early 1930s
and  
%
the contemporary discovery of the He I 1083 nm infrared line  in the solar spectrum by Harold and Horace Babcock \citep{Babcock_1934}. Since then several coronal quantities have been measured daily. These quantities have been  supplemented by the huge data set of imaging observations taken during the space era. These data 
allow to evince the magnetic structure of the corona that originates the solar wind governing the   magnetic field in the heliosphere. Direct measurements of the coronal magnetic field are still scarce to date. However, the coronal  field can be modeled by extrapolating the magnetic field observed in the inner solar atmosphere. 
To this purpose, 
sunspot data series have been employed as  input data to e.g. solar surface transport models coupled with extrapolations of the heliospheric field 
\citep[e.g.][and references therein]{Jiang_etal2011}. 
 }

The He I line is seen in absorption on the solar disk, but in emission under some special conditions, e.g. during solar flares. The helium absorption   is enhanced  with respect to the quiet Sun above active regions and H$_{\alpha}$ filaments  \citep[e.g.][]{Brajsa_etal1996}, and reduced  in coronal holes  (CH)\footnote{Coronal holes are areas where the Sun's corona is darker, and colder, and has lower-density plasma than average. They are associated with rapidly expanding open magnetic fields and the acceleration of the high-speed solar wind. Find more information in e.g. \citet[][]{Potgieter_2013}.}. 
 Due to a complex
line-formation mechanism \citep[][]{Avrett_etal1994}, the
patterns seen in He I 1083 nm line observations are affected by processes in the upper 
chromosphere, transition region, and low corona. In particular, model calculations  show a strong dependence of the line absorption on coronal illumination. \citet[][]{Lagg_2007} summarizes  the results of  magnetic field measurements using the He I line and discusses the potential of this  line as a diagnostic of the solar  outer atmosphere. Observations at this line have been taken at the NSO Kitt Peak since 1974 on a daily basis.

{\bf The solar upper atmosphere also emits radio flux.
It consists of   free-free emission from quiet sun coronal plasma and gyromagnetic emission from sunspots in active regions \citep[e.g.][and references therein]{Shibasaki_etal2011}.
}

The integrated radio flux from the Sun is defined as:
\[
F=\frac{2k_{B} }{\lambda^{2}}\int {T_{b} } d\Omega ,
\]
where, $F,\mbox{\, }k_{B} ,\mbox{\, }\lambda \mbox{,\, }\Omega $, are radio 
flux (Wm$^{-2}$Hz$^{-1})$, the Boltzmann constant, the observing wavelength, 
and the solid angle of the source respectively. The solar radio flux 
is measured using the Solar Flux Unit {\bf (SFU, 10$^{-22}$ 
Wm$^{-2}$Hz$^{-1})$}. $T_{b} $ is called radio brightness temperature and is 
related to the plasma temperature $T$ as follows:
\[
T_{b} =\int {Te^{-\tau }} d\tau ,
\]
where, $\tau $ is the optical depth, which  is related to the absorption 
coefficient $\kappa$ by the relation $\quad d\tau =\kappa d\ell $, where $\ell $ is the path 
length along the line of sight. In the case of thermal free-free emission 
(collisions of electrons with ions), the absorption coefficient is a 
function of the plasma density $n$, the temperature, and the observing 
wavelength: $\kappa =\xi c^{-2}\lambda^{2}n^{2}T{ }^{-3/2}$, where $\xi $ 
is a constant (0.1 in the chromosphere and 0.2 in the corona) and $c$ is the 
light speed. In an optically thick case $(\tau \gg 1)$, the brightness 
temperature is equal to the plasma temperature and the integrated flux is 
inversely proportional to square of the measured wavelength. 
Gyro-resonance 
is prominent around 3 GHz compared to free-free contribution.


 
Among the various radio measurements, those pertaining the  flux  in the wavelength range of  2.8 GHz  or 10.7 cm, 
 near the peak of the observed solar radio emission, constitute  the longest, most stable and almost uninterrupted record of direct physical data of solar activity available to date. This is because the radio flux measurements are rather insensitive to weather and disturbances in the ionosphere. Historically, this index, which provides the so-called {\bf F10.7 index series,  has been used as an input to ionospheric models as a surrogate for the solar UV output  that produces photoionization in the Earth's ionosphere.}

\subsection{He 1083 nm line observations}

Regular observations of the solar disk at the He I line started in late 1930s.  These observations  were carried out with spectroheliographs  until  the availability of spectropolarimetric measurements.   {\bf The results were mostly employed to investigate  the chromospheric topology and dynamics.}  After their first detection from space in the late 1960s and early 1970s, most of CH data  available to date were later deduced from available He I 1083 nm  observations. 

\begin{figure*}
  \includegraphics[trim=2cm 2cm 2cm 2cm,clip=true,width=8cm,angle=270]{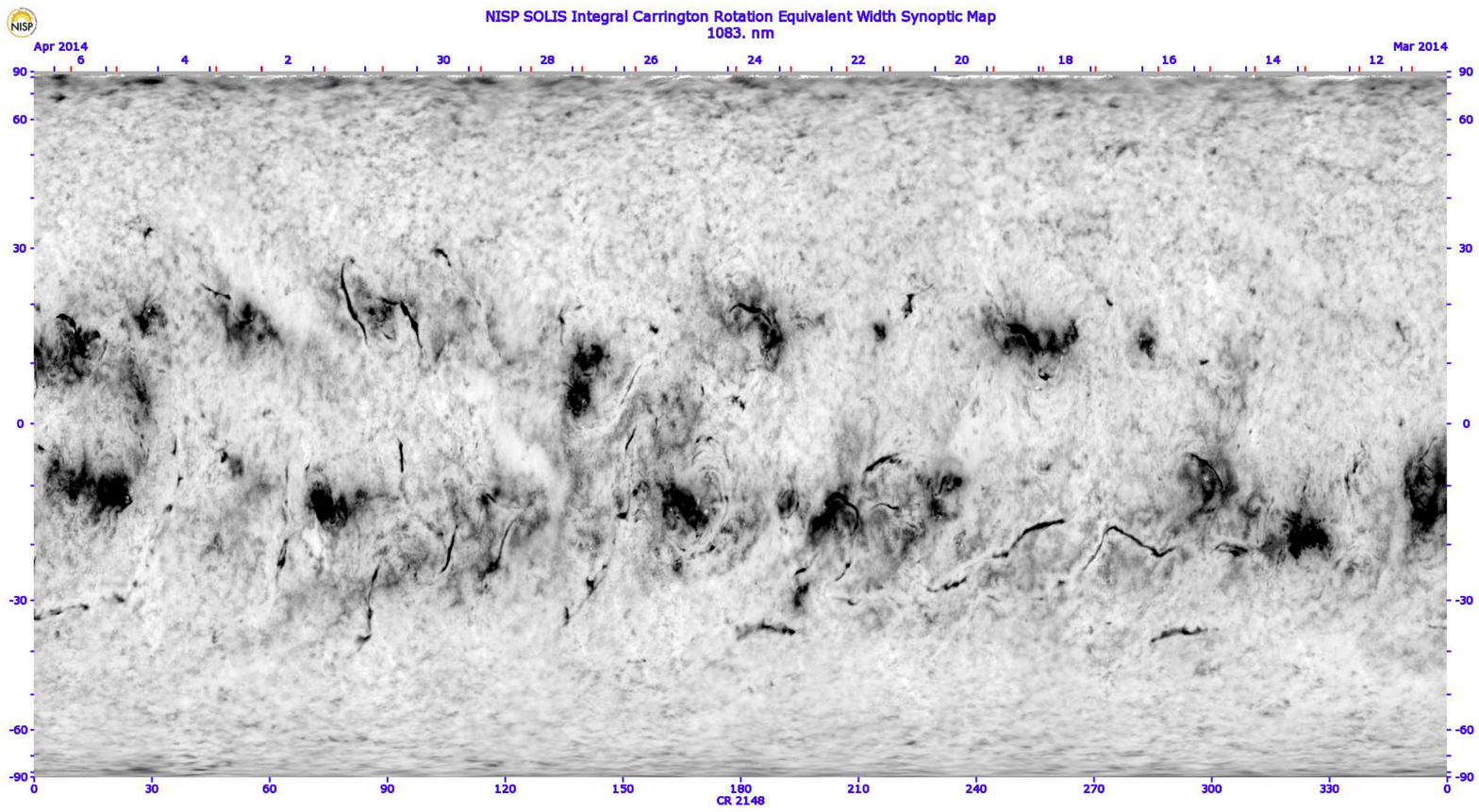}
\caption{Map derived from daily full-surface intensity values of the equivalent width as measured from profiles of the spectral line of He I 1083.0 nm. These measurements are from the NSO archive. }
\label{fig:7}       
\end{figure*}

Full-disk observations at the He I 1083 nm line have been taken at the NSO Kitt Peak since 1974 \citep[][]{Harvey_etal1977} on a daily basis, {\bf by using } the KPVT magnetograph. The data  were published by the  NOAA/NGDC archive\footnote{http://www.ngdc.noaa.gov} as Helium synoptic charts for each solar rotation or as CH contours plotted on H$_{\alpha}$ 
synoptic charts.
In particular, the latter method was used to determine the polarity of 
each observed CH, and to identify it over several solar 
rotations.  Magnetograms from NSO Kitt Peak  were also 
used for this purpose.  
The NSO He I 1083 nm observations have been continued by the SOLIS VSM since 2003. The data obtained consist of  full-disk intensity maps representing the equivalent width as measured over the solar disk from profiles of the spectral line of He I 1083 nm.  

Figure \ref{fig:7} shows a SOLIS He I synoptic chart for the period from 12 March to 5 April 2014 from the NSO archive.
{\bf  The data in this archive  have been employed to derive information on CH positions.  
CHs were found  at the Sun's polar regions at solar minimum, and located anywhere on the Sun during solar maximum.  } Besides, transient ``dark points'' in He I 1083 nm observations were found to be associated with small magnetic bipoles. The number of these dark points resulted to vary inversely with the sunspot number \citep[][]{Harvey_1985}. 

The NSO Kitt Peak He I 1083 nm intensity {\bf maps   have been used e.g. by \citet[][]{Harvey_etal2002} to identify and measure the evolution of polar CHs during cycles 22 and 23. } They found that  polar CHs, which are the largest at cycle minimum,  evolve from high-latitude ($\approx$ 60$^{\circ}$) isolated holes. 
During the initial 1.2--1.4 years following the polar polarity reversal, the polar CHs develop asymmetric lobes extending to active latitudes {\bf and the area and magnetic flux of
the CH increase rapidly. }

\subsection{Radio flux data}

\begin{figure*}
{\includegraphics[trim=0.5cm 3cm 2cm 3cm,clip=true,width=8cm,angle=270]{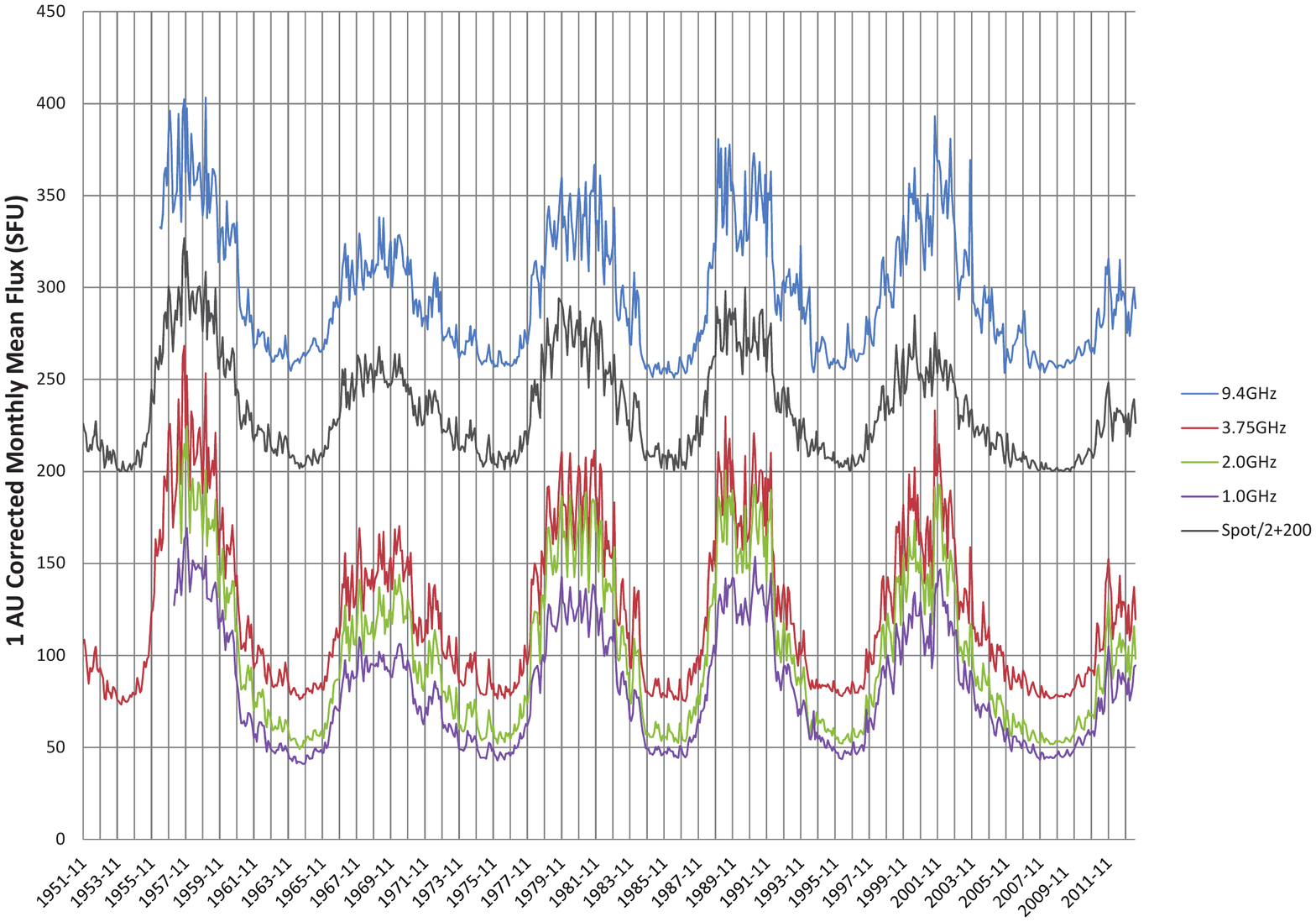}}
\caption{Monthly averaged values of the radio flux measurements at the 9.4, 3.75, 2, 1 GHz from the NSRO archive (January 1951 -- December 2013), corrected for 1 AU distance, superposed to scaled values of the ISN  from the {\bf ROB-SILSO} series. 
}
\label{fig:8}       
\end{figure*}


Synoptic radio measurements  have been made at the various observatories since 1945  \citep{Sullivan_1984}, 
by using  radio polarimeters, to measure the total coming flux and the circular-polarization degree  at various frequencies, and by employing radio heliographs made of many antennas  that allow to produce  interferometric measurements at various frequencies. 
The radio emission data available to date include daily values of solar flux density at 30 different frequencies ranging from 0.1  to 15
GHz, and tables of distinctive events. Radio charts of  active regions on the Sun have been also derived from the measurements carried out at e.g. the  Nancay and Nobeyama observatories. 

{\bf Radio flux from the Sun consists of three components  that can be 
discriminated by the temporal scale of flux variations \citep[e.g.][and references therein]{Tapping_2013}. 
The burst component  is 
the flux variation associated with  flares on  typical time scales from 
seconds to hours. 
It is employed to diagnose high 
energy particle acceleration and production of thermal plasma associated 
with  flares. The slowly varying (S) component is linked  with the evolution of active 
regions on typical time scales from a day to weeks. The 
base (B) component  is the residual flux  after subtraction of the  
burst and S components. Daily quiet time total flux (daily flux) data consists of 
S and B components. }


As stated above, {\bf the time series of the radio flux measured at 2.8 GHz or 10.7 cm, the F10.7 index  constitutes} the longest record of a physical data of the solar cycle available to present. 
This index is a   measurement of the integrated emission at 10.7 cm wavelength from all sources present on the disk. 
Daily measurements at local noon of this flux have been made  by the National Research Council  (NRC) of Canada from 1947 to 1991 in Ottawa and thereafter  in Penticton. The flux values \citep{Covington_1969} are expressed in the SFU units  introduced above. 
The data, which  are available at the site of the NRC\footnote{http://www.spaceweather.gc.ca/solarflux/sx-5-eng.php} and NOAA/NGDC 
archive\footnote{http://www.ngdc.noaa.gov/stp/space\_weather/solar\_data/solar\_features/solar\_radio/noontime\_flux/}, include
daily F10.7 flux values, monthly averages, and 
rotational averages computed over a solar  rotation. 
The NOAA/NGDC archive  also contain radio measurements obtained at various frequencies e.g. since 1956 at the Astronomical Observatory of the Jagellonian University in Cracow,  since 
1966 with the USAF Radio Solar Telescope Network operated at various  observatories, and from 1962 to 1973 with the  Stanford Radio Astronomy Institute  telescope. 



The second longest record of solar radio measurements is from Japan \citep[][]{Tanaka_etal1973}. Measurement at 3.75 and 9.4 GHz  started there in 1951 and 1956, respectively; they were supplemented  with  measurements at 1.0 and 2.0 GHz  in 1957.
The measurements were carried out at the Toyokawa observatory  till 1994, and thereafter  continued at the  Nobeyama Solar Radio Observatory (NSRO). Results of these and of the earlier synoptic measurements carried out from 1951 to 1994 
are available at the NSRO  site\footnote{http://solar.nro.nao.ac.jp/norp/}. 

{\bf
Figure \ref{fig:8} shows the time series of monthly averaged values of the radio flux measurements  from the NSRO archive,  superposed to scaled values of the ISN  from the {\bf ROB-SILSO} series.  Pearson's correlation coefficients 
between the radio flux   and  ISN values are range between $\approx$ 0.94 and 0.98.}


Additional radio data include those obtained with the Radio Solar Telescope Network started by the Sagamore Hill Solar Radio Observatory in 1966, with stations at  Palehua/Kaena, Learmonth, and SanVito, whose observing frequencies are 245, 410, 610, 1415, 2695, 4995, 8800, and 15400 MHz,  and by the  
Kislovodsk station of the Pulkovo Observatory\footnote{http://www.gao.spb.ru/english/database/sd/tables.htm}, which started operation in 1960 at  wavelengths of 5, 10, 15 GHz (or 2, 3, 5 cm).
Besides, since 1996 
the Nancay Radio Heliograph \citep{Kerdraon_1997} and decametric array telescopes \footnote{http://secchirh.obspm.fr/nrh\_data.php} have been acquiring   thousands of interferometric images each day of the solar corona and at very  low frequency. The frequencies monitored regularly with these telescopes
cover the range  5 -- 10 Mhz, 10--100 MHz (from 20 to 75 MHz), higher than 100 MHz (at
164 MHz and 327 MHz). It is worth noting that  1D scans of the Sun were also taken from 1967 to  1996, but the data are not available online. Data from the decametric telescope  goes back to 1991 \footnote{http://realtime.obs-nancay.fr/dam/data\_dam\_affiche/}.


The daily F10.7 index was found to be well correlated to sunspot number and area \citep{Denisse_1949}. It was then noticed  that the radio emission in the 
range of 3 -- 30 cm wavelength, or 10 -- 1 GHz frequency range, correlates 
well with E-layer ionization \citep{Kundu_1970}. {\bf Since then the daily F10.7 index has 
been used as a proxy of solar cycle in models of the Earth's upper atmosphere.} 
Recent studies
 by \citet{Tapping_etal2011} and \citet{Svalgaard_etal2010} suggested that the relation between 
 sunspot numbers and radio flux has started 
to deviate from the previous relation since the 23 solar cycle. 
However, \citet{Henney_etal2012} found that the relation between total 
magnetic flux on the solar surface and radio flux values does not show any  
noticeable change in the last cycles. {\bf The results suggest that the F10.7 index is a better proxy of the total  magnetic flux on the 
solar surface than sunspot numbers.
}

In a recent study, \citet{Dudok_etal2014} merged daily observations from the Ottawa/Penticton and Toyokawa/Nobeyama observatories into a single homogeneous data set of the solar flux at wavelengths of  1, 2, 2.8, 3.75, 9.4 GHz (or 30,
15, 10.7, 8 and 3.2 cm), spanning from 1957 to present. They found  that most solar proxies, in particular the MgII index, are remarkably
well reconstructed by simple linear combination of radio fluxes at various wavelengths. {\bf Their results also indicate that the flux at 1 GHz (or 30 cm) stands out as an excellent
 proxy of solar cycle and is better suited than the F10.7 index for the modelling the Earth's thermosphere-ionosphere system. }

\subsection{Coronal indices}

\begin{figure*}
\includegraphics[trim=0.2cm 5cm 2cm 5cm,clip=true,width=10cm]{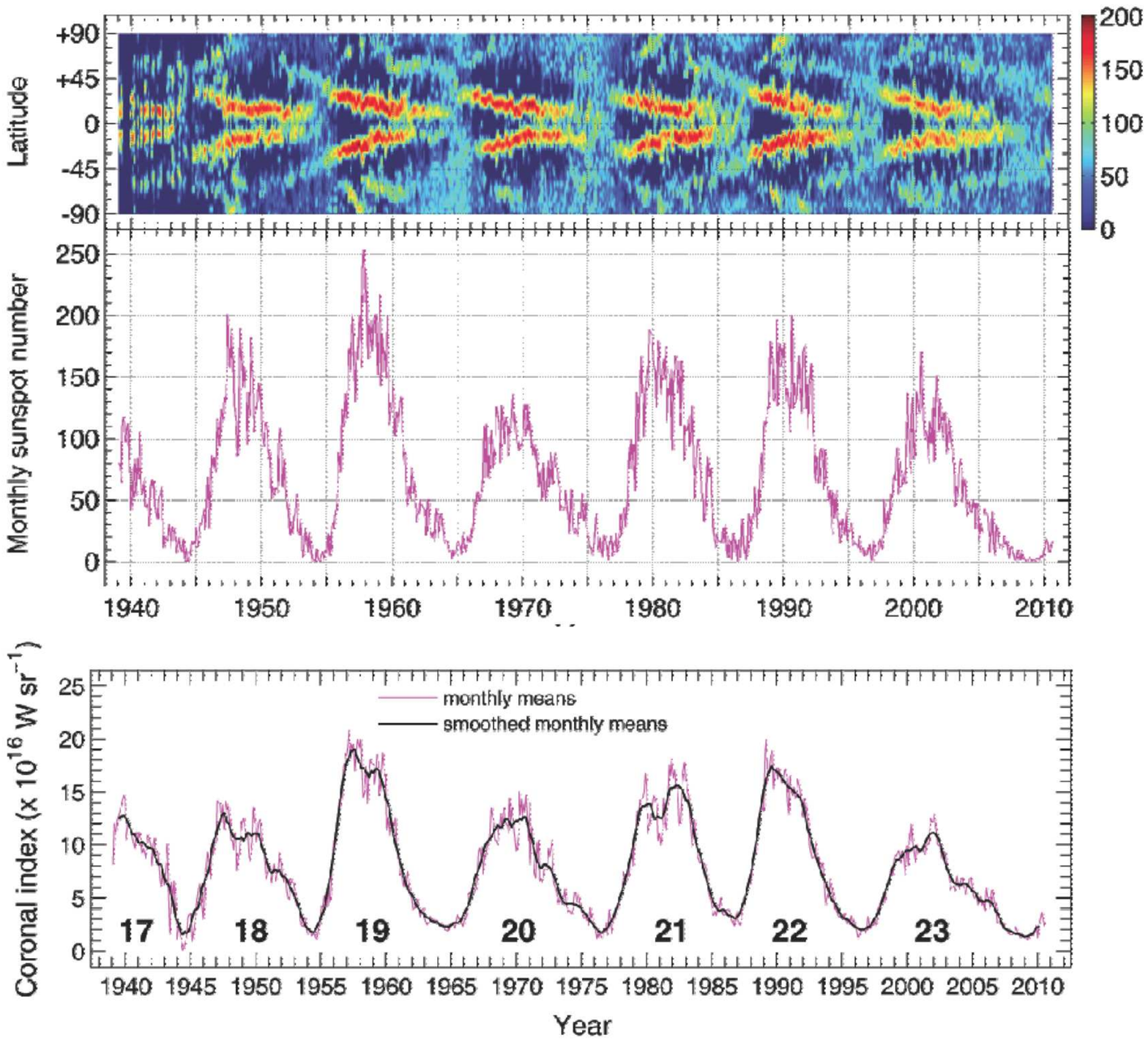}
\caption{
Time series of the coronal index (CI, bottom panel), the sunspot number (middle), and the Fe XIV 530.3 nm coronal green emission line brightness maxima (top). The CI is computed from the homogeneous dataset (HDS). Panels adapted from \citet{Minarovjech_2011}. 
}
\label{fig:10}       
\end{figure*}

The intensity of the  coronal green radiation from the Fe XIV emission line  at 530.3 nm  has been observed since 1939 at the Arosa observatory and then  recorded from 1947 onwards first at the  Climax and Pic du Midi observatories and thereafter at other sites (e.g. at Norikura Observatory\footnote{http://solarwww.mtk.nao.ac.jp/en/db\_gline1.html} between 1951--2009). The emission of this line can be observed above the entire solar limb during the whole solar cycle, contrary to the emission of other coronal lines, e.g. the Fe X red line at 637.4 nm and the 
Ca XV  yellow line at 569.4 nm, which can be  observed only occasionally. The intensity of
the green corona depends on both the density and temperature of the plasma in the outer solar atmosphere,
and both these quantities  are modulated by the local  magnetic fields. At present observations
of the green corona are made at the Kislovodsk,
Lomnicky Stit, and Sacramento Peak observatories. The measurements recorded at the various sites  differ due to time difference, method of observation, height of the observation
above the solar limb, and other parameters. However, the data obtained at these observatories have been processed since 1939 to obtain  a physical index of the solar cycle signature on the outer  solar atmosphere,  the so-called  coronal index \citep[CI, e.g.][and references therein]{Rybansky_etal2005}, and a
 homogeneous data set \citep[HDS,][see Figure  \ref{fig:10}]{Minarovjech_2011}.

This data set, which has been  obtained by scaling all the available measurements to the   photometric scale of the Lomnicky Stit observations, is available for the period of 1939 -- 2008 at the NOAA/NGDC  web page\footnote{http://www.ngdc.noaa.gov/stp}.
Recently, \citet{Dorotovic_2014} developed a method to substitute the ground-based observations by space-borne 28.4 nm (Fe XV) observations from the SOHO/EIT data. The dataset, named Modified Homogeneous Dataset (MHDS), extends the HDS beyond 2008. For the period up to 1996, the MHDS is identical with the former HDS. The MHDS is available online\footnote{http://www.suh.sk} 
for the 1996 -- 2010 period.

The coronal index represents the averaged daily irradiance in the green coronal line emitted in 1 steradian towards the Earth, and it is   expressed in (W sr$^{-1}$) units\footnote{Conversion: $1 \times 10^{16} W sr^{-1} = 4.5 \times 10^{-7}$ W m$^{-2}$ = 1.2 $\times$ photons cm$^{-2}$s$^{-1}$.}.  
  {\bf \citet[][]{Rusin_etal2002} analyzed these data to study } the relationship
between the intensity of the green corona and strength
of photospheric magnetic flux over the period 1976 -- 1999, by finding   a relation between these indices that allowed them to extend solar surface magnetic fields estimates since 1976 back to 1939, when the green corona began to be observed.
A discussion of the coronal index with respect to other solar cycle indices can be found in  e.g. \citet[][]{Rybansky_etal2001}.

In addition to the integrated flux measurements of the coronal green line emission, other measurements have been also recorded systematically, e.g. the intensity measured  around the solar limb,  within intervals of 5$^{\circ}$, at the  coronal lines at   637.4 nm (Fe X), 530.3 nm (Fe XIV) and 569.4 nm (Ca XV). These  lines are formed at approximate temperatures of 1, 2 and 3 MK, respectively. These measurements have been also processed to produce full-disk maps from 14 days of measurements projected onto a sphere.

\section{Conclusions}
For hundreds of years the evolution of the solar features driven by the cyclic variation of the solar magnetic field has been  monitored systematically. A variety of indices have been introduced in order to represent the many different results  derived from observations of the solar features  in time. These indices include values from measurements of  physical quantities, e.g. the Ca II K line emission and radio flux, as well as  of parameters derived from the observations, the most common being the sunspot numbers.
The time series of the sunspot  numbers, which  constitute 
one of the longest continuous measurement programs in the history of science,  continue to be used as the most common index to describe solar cycle properties. However, additional measurements of more physical quantities modulated by the solar cycle are also available online {\bf to enter  the models of e.g. the solar dynamo and of the Earth's climate response to solar variability}, though  the time series of the physical quantities are shorter than that of sunspot numbers.  

Continuation of the time series of the many indices of solar cycle available to date in the next decades is especially important {\bf since the long-term
trends of solar activity have become a topic of great interest and research after the last activity minimum}. 
Besides, continuous improvement of the accuracy of earlier data and of the whole series, precise long-term data calibration, and  extension of the time series back in time, can also lead to a better  knowledge of the solar cycle and its effects on the whole heliosphere. To this regard, the 
recovery, digitization, and analysis of historical observations started in recent years, consisting of both  
full-disk drawings dating back early 17th century and photographic observations  taken since 1876 at several observatories, promise to extract from these unexploited data 
 far more detailed information on solar magnetism than just the sunspot number and area records available till recent times. 
 
 The 
availability of long and accurate time series of solar cycle indices  
 has a  
{\bf large potential impact not only on  solar research but also on space weather and space climate  studies.  Indeed, sunspot number, UV and radio flux series are the key input data  to studies of  the solar activity impact on the Earth's upper atmosphere and ionosphere  \citep[][]{Maruyama_2010}, through e.g. multiscale models of traveling ionospheric disturbances \citep[][]{Fedorenko_etal2013} and data-driven analysis  \citep[][]{Kutiev_etal2013,Scott_etal2014}. Besides, sunspot and plage area, Mg II and F10.7 measurement series enter models of total and spectral solar irradiance variations  on  long time scales \citep[e.g.][]{Krivova_etal2010,Lean_etal2011}. These models have been recently employed to study e.g. 
the influence of spectral solar irradiance variations on stratospheric heating rates  \citep[][]{Oberlander_etal2012,Anet_etal2013,Thuillier_etal2014}, to estimate 
the tendency of the Northern Hemisphere temperature for the next decades by using a thermodynamic climate model \citep{Mendoza_etal2010}, to describe the  long-term trends in the North Atlantic Oscillation Index  \citep[][]{vanLoon_etal2012},
and  to carry out chemistry climate model experiments of the impact of a time-varying  solar cycle and  quasi-biennial oscillation  forcings on the Earth's atmosphere and the ocean  \citep{Petrick_etal2012,Matthes_etal2013}. 

Finally, the availability  of long and accurate time series of solar cycle indices} may also improve our ability to predict the future evolution of the solar activity.

\begin{acknowledgements}
The authors are grateful to the International Space Science Institute, Bern, for the organization of the workshop ``The Solar Activity Cycle: Physical Causes and Consequences'', the invitation to contribute to it, and the kind support received to the purpose. The authors thank Fabrizio Giorgi  for preparing Figs. 1 to 8.
This study received funding from the European UnionÕs Seventh Programme for Research, Technological Development
and Demonstration,  under the Grant Agreements
of the eHEROES (n 284461, www.eheroes.eu),  SOLARNET (n 312495, www.solarnet-east.eu), and SOLID (n 313188,  projects.pmodwrc.ch/solid/) projects. It was also supported by
COST Action ES1005 ÔÔTOSCAÕÕ (www.tosca-cost.eu). LvDG's work was supported by the
Hungarian Research grants OTKA K-081421 and K-109276, and by the STFC Conso\-lida\-ted
Grant ST/\-H002\-60/1.

Final acknowledgements go to the many observers  and  astronomers, both amateur and professional, 
for performing the regular observations of the solar atmosphere and creating the databases of solar indices described in this paper. 

\end{acknowledgements}
\bibliographystyle{aps-nameyear}      
\bibliography{issi_papr_rev}                
\nocite{*}

%
%


\end{document}